\documentclass[useAMS,usenatbib,letterpaper]{mn2e}

\usepackage{graphicx}
\usepackage{fixltx2e}
\usepackage{url}

\usepackage{amsmath}
\usepackage{amssymb}

\newcommand{\acknowledgments}{\begin{small}
    \section*{Acknowledgments}\end{small}}
\newcommand\altaffilmark[1]{$^{#1}$}
\newcommand\altaffiltext[1]{$^{#1}$}
\newcommand{\gtsimeq}{\raisebox{-0.6ex}{$\,\stackrel
        {\raisebox{-.2ex}{$\textstyle >$}}{\sim}\,$}}

\title[Rayleigh-Taylor Stable and Unstable Contact Discontinuities with Anisotropic Conduction]{The Dynamics of Rayleigh-Taylor Stable and Unstable Contact Discontinuities with Anisotropic Thermal Conduction}

\author[Lecoanet, Parrish, and Quataert]{
\parbox[t]{\textwidth}{
Daniel Lecoanet,\altaffilmark{1}\thanks{E-mail:dlecoanet@berkeley.edu}
Ian J. Parrish,\altaffilmark{1} \&
Eliot Quataert\altaffilmark{1}}
\vspace*{6pt} \\
\altaffiltext{1}{Department of Astronomy and Theoretical Astrophysics
  Center, University of California Berkeley, Berkeley, CA 94720} 
}

\date{Sumitted to MNRAS 2012 February 8}
\begin{document}
\maketitle
\begin{abstract}
We study the effects of anisotropic thermal conduction along magnetic
field lines on an accelerated contact discontinuity in a weakly
collisional plasma.  We first perform a linear stability analysis
similar to that used to derive the Rayleigh-Taylor instability (RTI)
dispersion relation.  We find that anisotropic conduction is only
important for compressible modes, as incompressible modes are
isothermal.  Modes grow faster in the presence of anisotropic
conduction, but growth rates do not change by more than a factor of
order unity.  We next run fully non-linear numerical simulations of a
contact discontinuity with anisotropic conduction.  The non-linear
evolution can be thought of as a superposition of three physical
effects: temperature diffusion due to vertical conduction, the RTI,
and the heat flux driven buoyancy instability (HBI).  In simulations
with RTI-stable contact discontinuities, the temperature discontinuity
spreads due to vertical heat conduction.  This occurs even for
initially horizontal magnetic fields due to the initial vertical
velocity perturbation and numerical mixing across the interface.  The
HBI slows this temperature diffusion by reorienting
initially vertical magnetic field lines to a more horizontal geometry.  In
simulations with RTI-unstable contact discontinuities, the dynamics
are initially governed by temperature diffusion, but the RTI becomes
increasingly important at late times.  We discuss the possible application
of these results to supernova remnants, solar prominences, and cold
fronts in galaxy clusters.
\end{abstract}

\section{Introduction}

The interface between fluids of different densities can be
destabilized by gravity or other sources of acceleration.  If the
fluid is accelerated in a direction opposite of $\nabla\rho$, the
interface is unstable to the Rayleigh-Taylor instability (RTI).  When
a heavy fluid is on top of a lighter fluid, the RTI is characterized
by bubbles of light fluid rising into the heavy fluid, and downward
spikes of heavy fluid penetrating into the light
fluid \citep[e.g.,][]{chandra61}.  The RTI is potentially important in
a variety of astrophysical situations, e.g., in supernova remnants at
the contact discontinuity between the shocked supernova gas and the
shocked interstellar material \citep{gull73,che92,jn96}; in emerging
magnetic flux in the solar corona \citep{iso05,ber11}; and in buoyant
bubbles of gas produced by AGN in galaxy
clusters \citep[e.g.,][]{rob04}.

In some of these contexts, the plasma is hot and dilute.  If the
electron gyroradius is much smaller than the electron mean free path,
then the heat conduction by electrons is primarily along the magnetic
field \citep{bra65}.  We will refer to this effect as anisotropic
conduction.  Work by, e.g., \citet{bal00}, \citet{qua08},
and \citet{mcc11}, has shown that convection is strongly affected by
anisotropic conduction.  The normal Schwarzschild condition for
convective instability, $ds/dz<0$, is modified when the conductivity
time-scale is much shorter than the buoyancy time-scale.  In this `fast
conduction' limit, plasmas are unstable to a heat flux driven
buoyancy instability (HBI) when $\boldsymbol{g}\boldsymbol{\cdot}\nabla T<0$ or are
unstable to the magnetothermal instability (MTI) when
$\boldsymbol{g}\boldsymbol{\cdot}\nabla T>0$.  Although the HBI acts to reduce the
effective conductivity in the direction parallel to gravity, the MTI
produces vigorous convection \citep[e.g.,][]{ps07,pq08,mcc11}.

Our goal in this paper is to investigate the magnetized RTI when
anisotropic conductivity is important.  \citet{bal08b}
and \citet{bal08a} have simulated supernova remnants with anisotropic
conduction.  However, their simulations are of an entire supernova
remnant, and their analysis did not describe how the contact
discontinuity is affected by the combination of the RTI and
anisotropic conductivity.  In fact, in
\citet{tb06}, there does not seem to be any RTI type
perturbation of the contact discontinuity, unlike in, e.g., \citet{che92}.

Our analysis will make several simplifying assumptions.  We will focus
our attention only on weak magnetic fields that are not dynamically
important.  \citet{sg07} have shown that stronger magnetic fields
prevent fluid motions which bend field lines, and that in the strong
magnetic field limit, the RTI dynamics are dominated by interchange
motions.  Our assumption will be that the fields are strong enough to
keep the conduction anisotropic, but not so large that they are
dynamically important.  We will also largely (though not entirely)
neglect the effects of anisotropic viscosity, which acts to
preferentially damp motion along magnetic fields.  Previous work has
shown that this also leads to the RTI being dominated by interchange
motions \citep{ds09}.

In addition to studying the RTI, we will also investigate the effects
of anisotropic conduction on contact discontinuities that are RTI
stable.  \citet{bir10} propose that cold fronts in galaxy clusters can
form through the merger of shocks.  These observed cold fronts are
sharp contact discontinuities, and the intracluster medium (ICM) is
sufficiently hot and dilute for anisotropic conduction to be
important.  These contact discontinuities are likely to be RTI stable,
given their small widths.  \citet{pq08} and \citet{bir10} suggest that
anisotropic conduction could help to maintain the small widths of cold
fronts in clusters, because the HBI acts to suppress heat conduction
at RTI-stable contact discontinuities.

The remainder of the paper is organized as follows.  First, we perform
a linear stability analysis including the effects of anisotropic
conductivity on an RTI-unstable contact discontinuity with a weak
horizontal magnetic field (\S \ref{sec:review}).  Next, we discuss our
numerical methods for simulating this setup using fully non-linear
compressible MHD simulations in \S \ref{sec:numerics}.
In \S \ref{sec:RTInocond} we consider the hydrodynamic problem, i.e.,
without magnetic fields or anisotropic conduction.
Section \ref{sec:cond} studies RTI-stable and RTI-unstable
contact discontinuities, for a range of initial magnetic field orientations.
Finally, in \S \ref{sec:conclusion} we conclude and discuss the
astrophysical implications of this work.

\section{Linear Theory}\label{sec:review}

\subsection{Basic equations}\label{sec:basicequations}

Throughout this paper, we employ the same set of basic
equations.  They are as follows:
\begin{eqnarray}\label{eq:continuity}
\frac{d \rho}{dt}&=&-\rho\nabla\boldsymbol{\cdot}\boldsymbol{v}, \\
\rho \frac{d \boldsymbol{v}}{dt}&=& -\nabla
p+\frac{1}{4\pi}\nabla\boldsymbol{\times}\left(\nabla\boldsymbol{\times}\boldsymbol{B}\right)-g\rho\hat{\boldsymbol{z}}, \label{eq:momentum} \\
\frac{\partial\boldsymbol{B}}{\partial t}&=&\nabla\boldsymbol{\times}\left(\boldsymbol{v}\boldsymbol{\times}\boldsymbol{B}\right), \label{eq:magnetic} \\
\rho T\frac{ds}{dt}&=&-\nabla\boldsymbol{\cdot}\boldsymbol{Q}, \label{eq:energy}
\end{eqnarray}
where $\rho$ is the density, $\boldsymbol{v}$ is the velocity, $p$ is the
pressure, $\boldsymbol{B}$ is the magnetic field, $g$ is the strength of
gravity, and $\hat{\boldsymbol{z}}$ is the unit vector in the $z$ direction.
$T$ is the temperature, $s$ is the entropy, and $\boldsymbol{Q}$ is the heat
flux given by
\begin{equation}\label{eq:heatflux}
\boldsymbol{Q}=-\hat{\boldsymbol{b}}\kappa\hat{\boldsymbol{b}}\boldsymbol{\cdot}\nabla T,
\end{equation}
where $\hat{\boldsymbol{b}}$ is the unit vector in the $\boldsymbol{B}$
direction, and $\kappa$ is the thermal conductivity.  We also use
$d/dt$ to denote the full time derivative, i.e.,
$d/dt=\partial/\partial t+\boldsymbol{v}\boldsymbol{\cdot}\nabla$.  We will employ the
thermal diffusion coefficient $\chi=T\kappa/P$ in lieu of $\kappa$.
For simplicity, we will approximate $\chi$ to be constant.  Throughout
this paper, we will take the medium to be an ideal gas with ratio of
specific heats $\gamma=5/3$, and use units where $k_b\mu m_p=1$, where
$k_b$ is the Boltzmann constant, $\mu$ is the mean molecular weight,
and $m_p$ is the proton mass.  In these units the equation of state is
$P=\rho T$, and $T$ has units of velocity squared.  We will use
$\delta$ to denote perturbed quantities below, i.e., $\delta\rho$ is
the Eulerian perturbation of the density.  In the linear problem
discussed here, we will assume that the magnetic field is sufficiently
weak that it is dynamically unimportant, and thus the Lorentz force
can be dropped from the momentum equation.  However, the Lorentz
force term is kept in our non-linear simulations.  In several
simulations we also include effects due to anisotropic viscosity in
the momentum and energy equations (see \S \ref{sec:anisovisc}), but we
do not include anisotropic viscosity in our linear analysis.

\subsection{Background}

We will now describe the background quantities in our linear stability
problem and the properties of the perturbations.  In the RTI problem,
two layers of different densities and temperatures are separated by an
interface at $z=0$.  The pressure is taken to be continuous at the
interface.  We will perturb this equilibrium and investigate the
linear stability of these perturbations.  We restrict our attention to
magnetic fields which are initially horizontal.  This allows us to use
a background state with a density discontinuity (and thus a
temperature discontinuity) which still satisfies the energy equation
(eqn. \ref{eq:energy}).  To avoid any conduction within the upper and
lower layers, we will assume that the background state in each layer
is isothermal.  This means that the background pressure and density
vary as $\exp (-z/H)$ where $H$ is the density scale height.  $H$ is
related to $c_s$, the adiabatic sound speed, by $H=c_s^2/(\gamma g)$.
By using isothermal domains above and below the temperature
discontinuity, we can restrict our attention to instabilities caused
by the discontinuity, as opposed to conduction-mediated instabilities
within the upper and lower domains (e.g., the MTI or HBI).  Although
$\rho$ is no longer constant in the upper and lower isothermal layers,
we will still refer to $\rho_+=\rho|_{z=+\epsilon}$ and
$\rho_-=\rho|_{z=-\epsilon}$.

\begin{figure*}
\centerline{\hbox{\includegraphics[width=12.5cm]{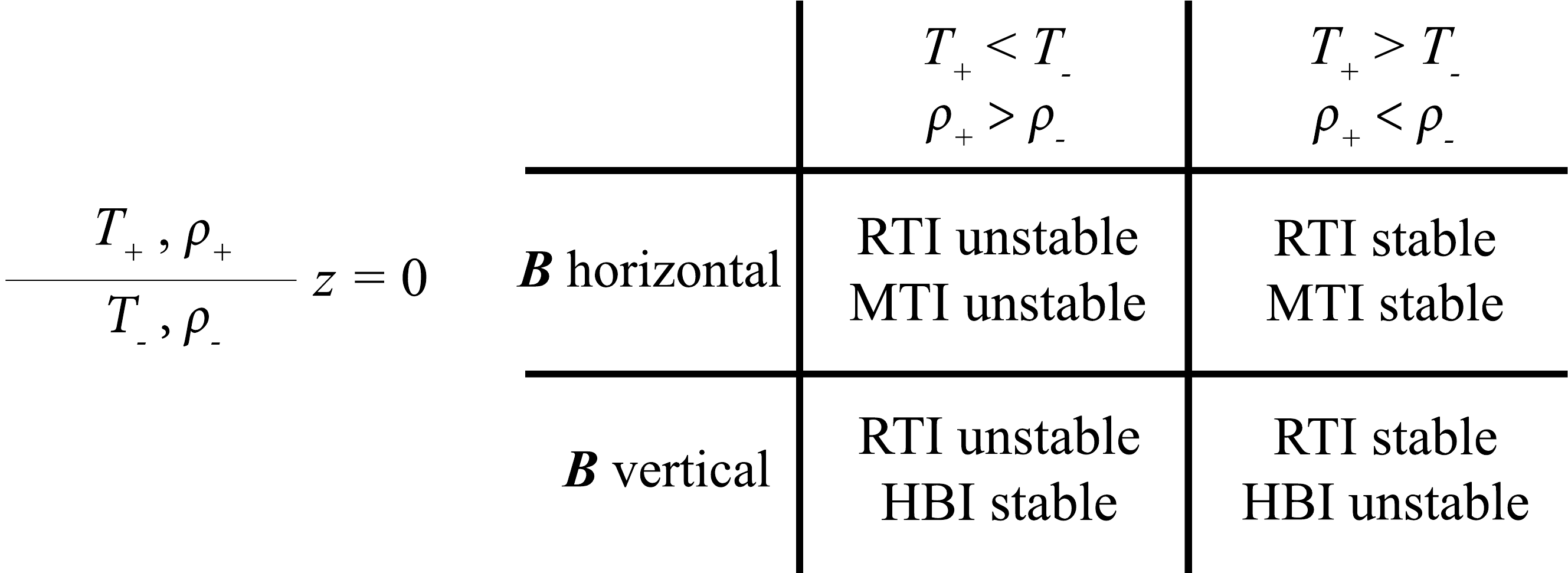}}}
\caption{A schematic of the various instabilities of a contact discontinuity with anisotropic thermal conduction.
  On the left we show the contact discontinuity with a temperature and
  density of $T_+$ and $\rho_+$ immediately above the interface, and a
  temperature and density of $T_-$ and $\rho_-$ immediately below the
  interface.  On the right we summarize the instabilities associated
  with each temperature and density jump, for either horizontal or
  vertical initial magnetic fields.}
\label{fig:chart}
\end{figure*}

The pressure is continuous at $z=0$ so the pressure jump,
$\Delta \left[P\right]$, is zero, where we define
\begin{equation}
\Delta\left[ f\right]\equiv f|_{z=+\epsilon}-f|_{z=-\epsilon},
\end{equation}
for any function $f$.  This implies $\Delta
\left[\rho\right]=-\Delta \left[T\right]$.  The RTI
corresponds to $\Delta\left[\rho\right]>0$, which yields
$\Delta\left[T\right]<0$, which we would expect to also be unstable to
the MTI \citep{bal00}.  However, states with
$\Delta \left[\rho\right]<0$, and thus $\Delta \left[T\right]>0$, are
RTI stable and we would expect to be MTI-stable as well (but for
non-horizontal fields, potentially HBI-unstable).  We summarize the
instabilities associated with each temperature and density jump, for
different initial magnetic field geometries, in Fig. \ref{fig:chart}.

We assume that all perturbation quantities vary in the horizontal and
time directions as
\begin{equation}\label{eq:fourier}
\delta\rho (x,y,z,t)=\delta\rho (z) \exp\left(i k_x x+i k_y y -i\omega t\right).
\end{equation}
Although we Fourier decompose the perturbations in the horizontal
directions, we do not Fourier decompose the perturbations in the
vertical direction.  We will denote by $k_\perp=\sqrt{k_x^2+k_y^2}$
the wavenumber perpendicular to gravity (\textit{not} perpendicular to
the magnetic field).

\subsection{Boussinesq limit}

Following e.g., \citet{chandra61} or \citet{shi08}, we will derive the
dispersion relation in two steps.  First, we use the
equations for the perturbations in the upper and lower layers to
derive the vertical structure of the perturbations for $z\neq 0$.
Second, we apply the boundary conditions.  The first three are that
the vertical kinetic energy decays to zero at $z\rightarrow\pm\infty$,
and that the vertical velocity, $\delta v_z$ is continuous at $z=0$.
The vertical derivative of the vertical velocity, $\partial_z\delta
v_z$, is in general discontinuous at $z=0$.  We thus take the limit of
the equations as $z\rightarrow 0$ to derive a jump condition
for $\partial_z\delta v_z$, which will lead to a dispersion relation.

First we will simultaneously employ the local and Boussinesq limits.
The local limit amounts to neglecting any derivative of background
quantities in the upper and lower domains in comparison to the
derivative of the perturbations ($H k_\perp\gg 1$).  In the Boussinesq
limit, $\delta P/P$ is small in comparison to e.g., $\delta\rho/\rho$.
In these limits, the momentum equation (eqn. \ref{eq:momentum}) gives
\begin{equation}\label{eq:RTvz}
\omega^2\nabla^2\left(\rho\delta v_z\right)-\omega^2\partial_z\nabla\boldsymbol{\cdot}\left(\rho\delta\boldsymbol{v}\right)=k_\perp^2g i\omega\delta\rho,
\end{equation}

First consider the upper and lower layers, where $z\neq 0$.  One can
use either the energy equation or the continuity equation to solve for
$\delta\rho$.  The energy equation implies
\begin{equation}
\left(i\omega-\omega_c\right)\frac{\delta\rho}{\rho}+\delta v_z\partial_z s=0,
\end{equation}
where
\begin{equation}
\omega_c=\frac{2}{5}\chi k_x^2
\end{equation}
is the conduction time-scale, and we take the magnetic field to be in
the $\hat{\boldsymbol{x}}$ direction.  The $\partial_z s$ term is
proportional to $1/H$, which suggests it might be small in the local
limit.  It can be shown \textit{a posteriori} that it is consistent to
ignore this term (using that $Hk_\perp\gg 1$).  When we assume that
$\delta v_z\partial_z s$ is small, we get that $\delta\rho = 0$ in the
upper and lower layers.  Then Equation \ref{eq:RTvz} reduces to
\begin{equation}
\left(-k_\perp^2+\partial_z^2\right)\delta v_z=0.
\end{equation}
This implies
\begin{equation}
\delta v_z\sim A_+ \exp\left(+ k_\perp z\right)+ A_-\exp\left(-k_\perp z\right),
\end{equation}
for some amplitudes $A_+$ and $A_-$.  This is the same result as for the
classical RTI analysis (i.e., constant density layers with no magnetic
fields or conduction), so anisotropic conduction has not affected the
vertical structure of the perturbations.

The next step is to apply the boundary conditions.  The boundary
conditions at infinity and the continuity of $\delta v_z$ at $z=0$
imply
\begin{equation}\label{eq:vertstructure}
\delta v_z(z)=\delta v_z(0)\exp\left(-k_\perp|z|\right),
\end{equation}
where $\delta v_z(0)$ is an amplitude.  In general we can write $\delta
v_z\sim\exp(k_{\pm}z)$, where the plus sign is taken for $z>0$ and the
minus sign is taken for $z<0$.  The boundary condition at infinity
requires $\mathrm{Re}(k_+)<1/(2H)$ and $\mathrm{Re}(k_-)>1/(2H)$.  In this
case, as for the classical RTI problem, we have $k_\pm=\mp k_\perp$.

Now we will discuss the jump condition on
$\partial_z\delta v_z$ at $z=0$.  Again we must combine the momentum
equation, Equation \ref{eq:RTvz}, with the energy equation.  The energy
equation, in the limit that $z\rightarrow 0$, can be written as
\begin{equation}
\left( i\omega-\omega_c\right)\left(-i\omega\delta\rho+\delta v_z\partial_z\rho\right)=0.
\end{equation}
Thus, we have that
\begin{equation}
-i\omega\delta\rho+\delta v_z\partial_z\rho=0
\end{equation}
regardless of the value of $\omega_c$.  This can be combined with
Equations \ref{eq:vertstructure} and \ref{eq:momentum} to give the same
dispersion relation as for the classical RTI problem,
\begin{equation}\label{eq:RTIdisprelation}
\omega^2=\frac{gk_\perp\left(\rho_--\rho_+\right)}{\rho_++\rho_-}\equiv -Agk_\perp,
\end{equation}
where $A\equiv (\rho_+-\rho_-)/(\rho_++\rho_-)$ is the Atwood number.

The key implication of Equation \ref{eq:RTIdisprelation} is that in
the local, Boussinesq limit the dispersion relation for the RTI is
independent of the magnitude of the thermal conduction, provided that
the magnetic field is aligned with the contact discontinuity.  We will
now briefly discuss this result.  In the local limit, $d\log P/dz$ is
much smaller than $k_\perp$, and we always have that
$\Delta\left[P\right]=0$.  Furthermore, the Boussinesq approximation
implies that $\delta P/P\ll\delta\rho/\rho$.  These facts together
mean that the Lagrangian pressure perturbation, $P'/P$ is small
everywhere, where we use $f'$ to denote the Lagrangian perturbation of
$f$.  The equation of state implies that
\begin{equation}
\frac{P'}{P}=\frac{\rho'}{\rho}+\frac{T'}{T},
\end{equation}
but since $P'/P$ is small, we have that
\begin{equation}\label{eq:isothermal}
\frac{T'}{T}=-\frac{\rho'}{\rho}=0,
\end{equation}
where the last equality is due to the continuity equation
(eqn. \ref{eq:continuity}).  This implies that in the Boussinesq
approximation, perturbations are isothermal, and thus are unaffected
by conduction.

The assumption that $\Delta \left[P\right]=0$ at the interface is a
key part of the RTI problem.  However, the assumptions that $\delta
P/P$ is small and that $d\log P/dz$ is much smaller than $k_\perp$
result from the Boussinesq and local approximations.

\subsection{Fully compressible linear theory}

We will now relax the Boussinesq and local approximations and solve the
linear fully compressible RTI problem including anisotropic thermal
conduction.  The solutions of the fully compressible equations are not
isothermal, and thus are affected by anisotropic heat conduction.

As above, we must first use the evolution equations in the upper and
lower layers to solve for the vertical structure of the
perturbations.  After some algebra, one finds the following equations
for $k_\pm$:
\begin{eqnarray}
k_+= \frac{1}{2H_+}\pm \sqrt{\frac{1}{4H_+^2}+k_\perp^2-R\frac{\omega^2}{g H_+}-\left(1-R\right)\frac{gk_\perp^2}{H_+\omega^2}}, \label{eq:kp} \\
k_-= \frac{1}{2H_-}\pm \sqrt{\frac{1}{4H_-^2}+k_\perp^2-R\frac{\omega^2}{g H_-}-\left(1-R\right)\frac{gk_\perp^2}{H_-\omega^2}}, \label{eq:km}
\end{eqnarray}
where $H_\pm=T_\pm/g$ is the density scale height in
the upper and lower layers, and
\begin{equation}
R\equiv\frac{\frac{3}{5}i\omega-\omega_c}{i\omega-\omega_c}.
\end{equation}
The quantity $R$ represents the ratio of the response time-scale for
pressure perturbations to the response time-scale for density
perturbations.  If $\omega\gg\omega_c$, then $R=3/5$ and
perturbations are adiabatic.  If $\omega_c\gg\omega$, then $R=1$ and
perturbations are isothermal.

Note that there are two roots for each of $k_+$ and $k_-$ in
Equations \ref{eq:kp} \& \ref{eq:km}.  The physical solution has
perturbations with finite kinetic energy at infinity, i.e.,
$\mathrm{Re}(k_+)<1/(2H_+)$ and $\mathrm{Re}(k_-)>1/(2H_-)$.  Since
the two roots of $k_+$ sum to $1/(2H_+)$ and the two roots of $k_-$
sum to $1/(2H_-)$, at least one root for each of $k_\pm$ will be
physical \citep{cun11}.

By integrating the momentum equation from $z=-\epsilon$ to
$z=+\epsilon$ and dropping terms of order $\epsilon$, we derive the
following dispersion relation:
\begin{equation}\label{eq:RTfull}
0=g k_\perp^2 \Delta\left[\rho\right]
-\omega^2\Delta\left[\frac{\rho}{1-R\frac{\omega^2}{T
      k_\perp^2}}\left(k_z-\frac{R}{H}\right)\right],
\end{equation}
where $k_z|_{\pm\epsilon}=k_\pm$.  Along with Equations \ref{eq:kp} \&
\ref{eq:km}, this can be used to solve for $\omega$ as a function of the
various parameters of the problem -- $g,
k_\perp, \omega_c, \rho_\pm$ and $T_\pm$.

Note that for the fully compressible problem, the values of $k_\pm$
depend explicitly on $\omega$, so we cannot solve for the vertical
structure of the perturbations before solving for the growth
rate -- instead we must solve for $\omega$ and $k_\pm$ simultaneously.
Also, because the quantities inside the square roots in
Equations \ref{eq:kp} \& \ref{eq:km} are generally complex, we do not
know \textit{a priori} if the square root term will have positive or
negative real part.  Our procedure for finding physical modes is as
follows.  We first pick a root for each of $k_+$ and $k_-$ and then
search for an $\omega$ which satisfies the dispersion relation
(eqn. \ref{eq:RTfull}).  Then we check that for this $\omega$,
$\mathrm{Re}(k_+)<1/(2H_+)$ and $\mathrm{Re}(k_-)>1/(2H_-)$.  If these
conditions are met, then the mode is physical.

To reduce the dimensionality of our parameter space, we
nondimensionalize lengths and times by setting $k_\perp=2\pi$ (i.e.,
we set the perpendicular length scale to one) and $g=1$.  We also take
the smaller of $\rho_\pm$ to be $1$.  The condition that $T_\pm=g
H_\pm$ means that setting $T_\pm$ determines the density scale height,
as well as determining the sound speed through $c_{s\pm}^2=\gamma
T_\pm$.  Continuity of pressure at $z=0$ requires that
$T_+\rho_+=T_-\rho_-$.  With all this in mind, the problem has the
following degrees of freedom.  First, we must pick if the top or
bottom of the density discontinuity is at the lower density of
$\rho=1$, as well as the density on the other side of the
discontinuity (these together are equivalent to picking an Atwood
number).  Next, we pick the scale height on one of the sides, which
uniquely sets the scale height on the other side.  Finally, we must
pick $\omega_c$.

\begin{figure}
\includegraphics[width=0.45\textwidth]{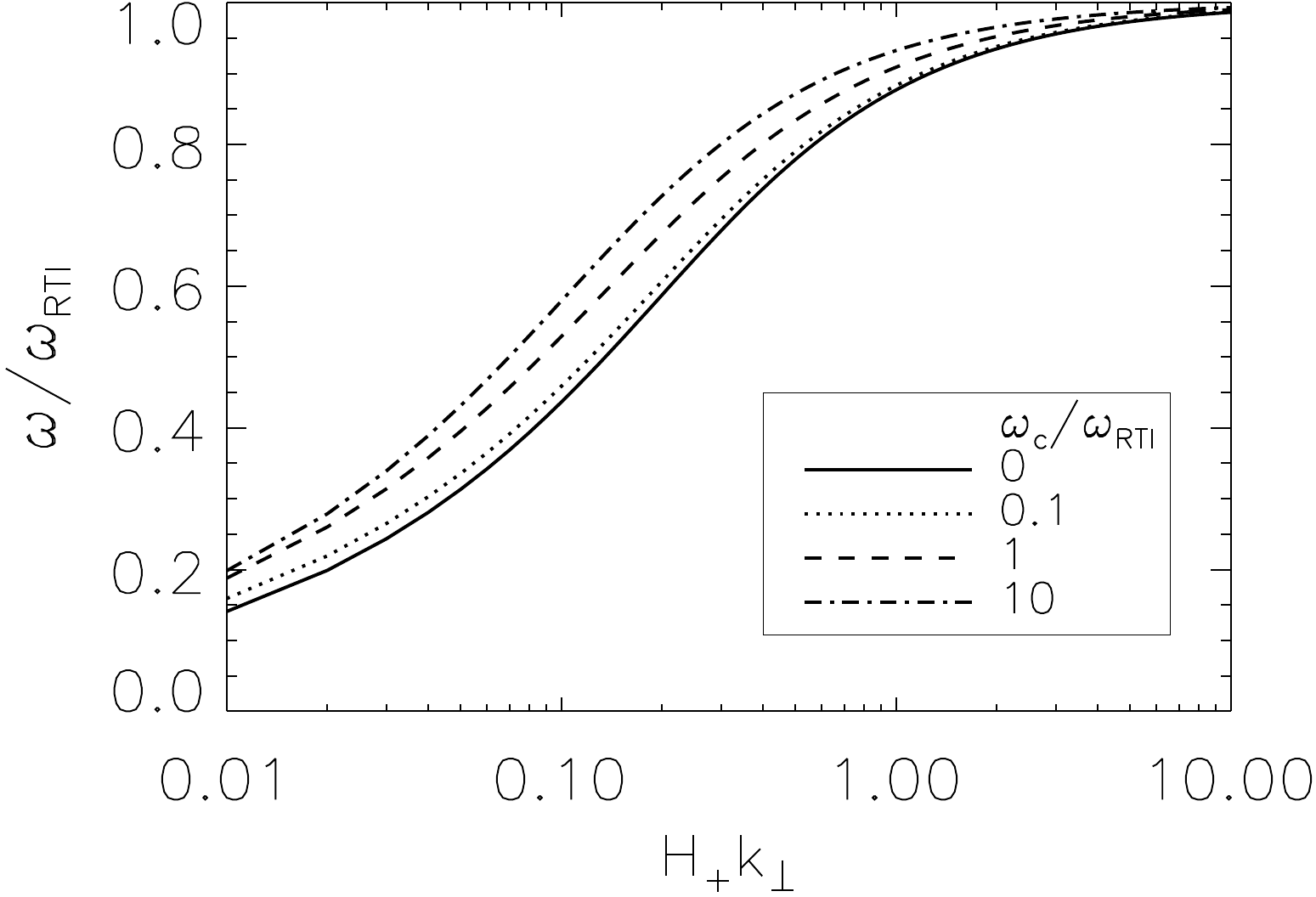}
\caption{Growth rates, $\omega$, normalized to
  $\omega_\mathrm{RTI}=\sqrt{Agk_\perp}$ as a function of $H_+
  k_\perp$ for various values of anisotropic thermal diffusion,
  parameterized by $\omega_c/\omega_{\mathrm{RTI}}$ (the ratio of the
  conduction frequency to the classical RTI growth rate).  For $\omega_c/\omega_{\mathrm{RTI}}>10$, we are
  in the isothermal, or fast conduction, limit and the results are
  nearly equivalent to those shown for
  $\omega_c/\omega_{\mathrm{RTI}}=10$.  As $H_+k_\perp$ decreases,
  compressibility becomes more important and
  $\omega/\omega_{\mathrm{RTI}}$ decreases.  At a fixed $H_+k_\perp$,
  the growth rate increases as $\omega_c$ increases from zero -- an
  asymptotic value is reached in the isothermal, or fast conduction,
  limit when $\omega_c/\omega_{\mathrm{RTI}}\gg 1$.}
\label{fig:omega}
\end{figure}

For the unstable case, we take $\rho_-=1$.  For simplicity, we will
take $\rho_+=3$ ($A=1/2$), but other values of $\rho_+$ give
qualitatively similar results.  We now have only two remaining
parameters: $H_+$ and $\omega_c$.  In Fig. \ref{fig:omega}, we plot
the growth rate, normalized to
$\omega_{\mathrm{RTI}}=\sqrt{Agk_\perp}$ as a function of $H_+k_\perp$
for several values of $\omega_c/\omega_{\mathrm{RTI}}$.  Without
compressible effects, the growth rate would be one when normalized to
$\omega_{\mathrm{RTI}}$.  Note that compressibility effects become
important as $c_s^2 k_\perp^2$ becomes the same size as $\omega^2$,
i.e., when $H_+ k_\perp\sim 1$.  Compressibility decreases the growth
rate of the instability, as energy is used to compress the fluid.  At
large $\omega_c$, the perturbations are isothermal (i.e., we are in
the fast conduction limit), whereas at small $\omega_c$, the
perturbations are adiabatic.

We will now discuss the differences between the large conductivity
(isothermal, fast conduction) and small conductivity (adiabatic)
limits.  To make the analysis easier we will take the limit in which
$H_+k_\perp$ is small.  This corresponds to the limit in which the
wavelength of the mode is much larger than the scale height of the
background.  It will turn out that
$\omega^2/\omega_{\mathrm{RTI}}^2\sim H_+k_\perp$ as
$H_+k_\perp\rightarrow 0$, which we will check \textit{a posteriori}.
First consider the isothermal limit.  When $R=1$, the equations for
the vertical structure (eqn. \ref{eq:kp} \& \ref{eq:km}) become much
simpler as the $1-R$ terms are zero.  Given that
$\omega^2/\omega_{\mathrm{RTI}}^2\sim H_+k_\perp$, the most important
term in the square root when $H_+k_\perp$ is small is the $1/4H_\pm^2$
term.  Thus, in this limit, $k_+\approx 0$ and $k_-\approx 1/H_-$.

The dispersion relation then becomes
\begin{equation}\label{eq:dispiso}
\frac{-\omega^2}{H_+}\left(\frac{\rho}{1-\frac{\omega^2}{T_+k_\perp^2}}\right)=gk_\perp^2\left(\rho_+-\rho_-\right).
\end{equation}
This can be easily solved to give
\begin{equation}\label{eq:omegaiso}
\omega^2=-gk_\perp^2 H_+\frac{\rho_+-\rho_-}{\rho_-},\hspace{0.5 in}\omega_c/\omega\gg 1, H_+k_\perp\ll 1.
\end{equation}
We see that $\omega^2$ differs from $\omega_{RTI}^2$ by a factor of
$H_+k_\perp (\rho_++\rho_-)/\rho_-$.

Now consider the small conductivity (adiabatic) limit.  We will still
take the limit in which $H_+k_\perp$ is small (i.e., the long wavelength limit).
This problem is more difficult as the $1-R$ terms in
Equation \ref{eq:kp} \& \ref{eq:km} are now important and $\omega$ appears
explicitly in the formula for $k_\pm$.  We find that
\begin{eqnarray}
H_+ k_+&=&\frac{1}{2}-\sqrt{\frac{1}{4}-\frac{2}{5}\frac{1}{\overline{\omega^2}}}, \label{eq:kzpad} \\
H_- k_-&=&\frac{1}{2}+\sqrt{\frac{1}{4}-\frac{2}{5}\frac{1}{\overline{\omega^2}}\frac{\rho_+}{\rho_-}}, \label{eq:kzmad}
\end{eqnarray}
where $\overline{\omega^2}=\omega^2/(H_+^2k_\perp^2)$ is a normalized growth
rate.  The dispersion relation can then be written as
\begin{eqnarray}
& &\overline{\omega^2}\left[\left(1-\frac{3}{5}\overline{\omega^2}\frac{\rho_-}{\rho_+}\right)\left(
  H_+k_+-\frac{3}{5}\right)\right. \nonumber \\
& &-\left.\left(1-\frac{3}{5}\overline{\omega^2}\right)\left(\frac{\rho_-}{\rho_+}\right)^2\left(H_-k_-
  - \frac{3}{5}\right)\right] \nonumber
\\ &=&\left(1-\frac{\rho_-}{\rho_+}\right)\left(1-\frac{3}{5}\overline{\omega^2}\frac{\rho_-}{\rho_+}\right)\left(1-\frac{3}{5}\overline{\omega^2}\right). \label{eq:dispad}
\end{eqnarray}
This is an equation for $\overline{\omega^2}$ with only one free
parameter, $\rho_-/\rho_+$.  We were unable to find a general analytic
solution for arbitrary values of $\rho_-/\rho_+$.  However, the
equation can easily be solved numerically.  For $\rho_-/\rho_+=1/3$ or
$A=1/2$, we have that $\overline{\omega^2}\approx -0.992$.  Note that
for the isothermal calculation, when $\rho_-/\rho_+=1/3$ or $A=1/2$ we
have $\overline{\omega^2}=-2$.

We generally find that the growth rate for isothermal perturbations is
faster than for adiabatic perturbations.  This can be explained by
describing how isothermal and adiabatic perturbations move in an
isothermal atmosphere.  The ratio of densities between a perturbed
fluid element and its surroundings remains constant if the
perturbation and background are isothermal.  However, for adiabatic
perturbations, the density of an upward propagating (low density)
perturbation drops less quickly than the density of the background, so
the perturbed element is less buoyant as it rises.  Similarly, an
adiabatically falling (high density) perturbation increases in density
less quickly than the background, so the perturbed element feels a
smaller downwards gravitational force as it falls.

Increasing the anisotropic conductivity from zero to infinity, the
value of the $R$ parameter moves from one to $3/5$ through the complex
plane.  We have not found any overstabilities in this problem.  The
behavior at low $H_+k_\perp$ is given by
Equations \ref{eq:kzpad}-\ref{eq:dispad}, replacing $2/5$ by $1-R$ and
$3/5$ by $R$.  The growth rate increases smoothly from the adiabatic
value to the isothermal value as $\omega_c$ increases.  Most of the
change in $\omega$ as $\omega_c$ varies occurs when $\omega_c$ is near
the adiabatic and isothermal growth rates.

The most prominent feature of the growth rates as $H_+k_\perp$ becomes
small (i.e., for long wavelength modes) is that the growth rates also
become very small.  One possible interpretation is that these modes
are very compressible and most of the potential energy gained by
changing vertical position is absorbed by compression.  A different
interpretation involves the background density profile.  Assume
that a mode of wavelength $k_\perp$ is `sensitive' to a distance
about $1/k_\perp$ above and below the interface.  If $H_+k_\perp\gg
1$, then the density is almost constant above and below the interface.
However, if $H_+k_\perp\ll 1$, there are many density scale heights
that can fit within a length $\sim 1/k_\perp$.  In this limit, the
density jump at $z=0$ looks like a small perturbation to an almost
isothermal atmosphere.  In this situation the RTI is relatively
insignificant because the density jump is small compared to the
isothermal stratification, so the growth rate should also be small.

\section{Numerical Methods}\label{sec:numerics}

We will now discuss the evolution of the RTI in the non-linear regime.
We will first discuss our numerical methods, and then review and
clarify previous results on the hydrodynamic RTI.  We then run
simulations of the RTI problem with conduction, with various magnetic
field geometries.  We will primarily be tracking the non-linear
evolution by measuring the height of the highest buoyant bubble as a
function of time.

We perform fully non-linear simulations using the Athena code
\citep{gs08,sto08}, which solves the compressible evolution
equations in conservative form using a Godunov method.  We implement
anisotropic conductivity as described in \citet{ps05}
and \citet{sh07}.

The simulation domain is Cartesian, with the computational domain
$x,y\in [-L/2,+L/2]$, $z\in [-L,+L]$ for some box length $L$.  We
normalize our lengths to $L$ (i.e., set $L=1$) in all our simulations.
Simulations are run at three resolutions: 64x64x128, 128x128x256, and
256x256x512.  The boundary conditions are periodic in the horizontal
directions.  At the top and bottom boundaries, the pressure is
extrapolated to maintain hydrostatic equilibrium, and reflecting
boundary conditions are applied to the remaining variables.

Our initial conditions for temperature are $T=T_+$ for $z>0$ and
$T=T_-$ for $z<0$.  The pressure and density, within the isothermal
domains, are given by
\begin{equation}\label{eq:backgroundPrho}
P_\pm(z),\rho_\pm(z)\sim \exp(-gz/T_\pm),
\end{equation}
so the scale height within the domains is $H_\pm=T_\pm/g$.  We will
refer to $\rho_\pm = \rho|_{\pm\epsilon}$.  In all of our simulations
we take the smaller of $\rho_\pm$ to be one and the larger to be
three.  The ratio $H_\pm/L$ relates the sound crossing time of the box
to the RTI growth time for perturbations with wavelengths comparable
to the size of the box.  In all of our simulations, we pick $H/L=100$
and $g=1$, where $H$ is the scale height on the $\rho=1$ side of the
interface.  We take $\gamma=5/3$, so $c_{s}^2=\gamma T=500/3$ on the
side of the interface with $\rho=1$.  The sound crossing time is
$\approx 13$, in comparison to the RTI growth time for modes the size
of the box, $\approx 1.7$.  Thus, our simulations are fairly
incompressible.

The uniform magnetic field is taken to be horizontal, vertical, or
forming a 45 degree angle with the horizontal.  The magnetic field
strength is set to $B/\sqrt{4\pi}=0.0001$.  For the linear RTI, such a
field would produce noticeable dynamical effects only on length scales
comparable to $5\times 10^{-7}$ \citep[hereafter SG07]{sg07}, which
cannot be resolved by our simulations.  Furthermore, the Alfv\'{e}n
velocity is about $10^{-5}c_s$.  Thus, the magnetic field plays no
dynamical role.

The conductivity, $\chi$, is normalized to $\sqrt{gL^3}$.  That
is, if we take $\omega_{RTI}=\sqrt{2\pi Ag/L}$ and
$\omega_c=\frac{2}{5}\chi 4\pi^2/L^2$, where $L$ is the length of the
box, then $\omega_c/\omega_{RTI}=2(2\pi)^{1.5}/(5\sqrt{A})\chi\approx
9\chi$ for our problem.  We will primarily pick values of $\chi$
smaller than one to understand the interaction of conductivity with
the RTI.

We start the simulations with a small perturbation to the vertical
velocity.  We set the velocity to $A_0R(1+\cos(\pi z/L))$, where $A_0$
the amplitude and $R$ is a random number between $-1$ and $+1$, as in
SG07.  The $z$ dependence is chosen so there is no vertical
perturbation at the top and bottom boundaries.  We take $A_0=0.05$,
which implies that the maximal velocity perturbations (of size
$2A_0=0.1$) are $\sim 0.01c_s$.  The initial perturbations are large
enough to start the simulations close to the non-linear regime.  In Appendix A we
discuss simulations with smaller initial perturbations which probe
the linear regime discussed in \S \ref{sec:review}.

\section{RTI Without Conductivity}\label{sec:RTInocond}

Before presenting the results of our simulations, we briefly discuss
previous results on the non-linear evolution of the RTI as presented in
\citet{dim04}.  Non-linear effects become important for
perturbations of wavelength $\lambda=2\pi/k$ when the perturbation has
moved the interface a height $\sim\lambda$ up or down.  The ascending
light fluid is described as a bubble, whereas the descending
heavy fluid is described as a spike.  The main diagnostic we
will use is the height of the highest bubble as a function of
time.  The results for the depth of the lowest spike as a function of
time are qualitatively similar.  In laboratory experiments and
numerical simulations, the height $h$ is reported to grow as
\begin{equation}\label{eq:height}
h=\alpha_h A g t^2,
\end{equation}
where $A$ is the Atwood number and $\alpha_h\approx$0.04--0.08 is a
dimensionless number arrived at experimentally.  Numerical simulations
produce values of $\alpha_h$ around a factor of two lower than
physical experiments -- this is thought to be due to increased
diffusion at the interface in numerical simulations \citep{dim04}.

\begin{table}
\caption{Parameters for simulations without conductivity.}
\label{tab:nocond}
\begin{center}
\begin{tabular}{@{}ccc}
\hline
Name & Resolution  & Constant $\rho$ or $T$ \\
\hline
SG & 256x256x512 & $\rho$ \\
RTLR & 64x64x128 & $T$ \\
RT & 128x128x256 & $T$ \\
RTHR & 256x256x512 & $T$ \\
\hline
\end{tabular}
\end{center}
\medskip
The simulations have either constant $\rho$ layers, as in SG07, or
constant $T$ layers.  The simulation with constant $\rho$ layers is
run using the same parameters as the hydrodynamic run in SG07.  These
parameters are described at the beginning of
\S \ref{sec:RTInocond}.  The simulations with constant $T$ layers
are run as described in \S \ref{sec:numerics}.
\end{table}

To provide some context for the calculations with anisotropic
conduction to follow, we now make a detailed comparison with the
hydrodynamic RTI simulation of SG07.  We have run a numerical
simulation using the exact same parameters as reported in
SG07.\footnotemark  We compare this to a simulation using the initial
conditions described in \S \ref{sec:numerics} that we will use for our
simulations with anisotropic thermal conduction.

\footnotetext{Specifically, we ran the problem with constant density layers with
$P=3/5$ at $z=0$ at a resolution of 256x256x512, without magnetic
fields or conduction, and took $L=g=0.1$.  There is some
ambiguity regarding the initial perturbation in SG07.  They state that
their initial perturbation is smoothed toward the vertical boundaries,
but then write that $v_z(z)=A_0R(1+\cos (2\pi z/L))$, where $A_0$ is
an amplitude and $R$ is a random number between $-1$ and $+1$.  Note
that $v_z$ is maximal at the boundaries if the vertical boundaries are
at $z=\pm L$.  We believe this is a typo and they actually set
$v_z(z)=A_0R(1+\cos(\pi z/L))$, so that $v_z(\pm L)=0$ -- we use this
functional form for the perturbation.  We also take $A_0=0.005$.}

In previous work on the RTI with constant density layers, the height
of the highest bubble is typically defined using the fraction of heavy
fluid ($f_h$) and the fraction of light fluid $(f_l=1-f_h)$ within a
volume \citep{dim04}.  If we use $\rho_+/\rho_-=3$, then
$\rho=2f_h+1$.  At $t=0$, $f_h=1$ above the temperature discontinuity.
Thus, many define the height of the highest bubble to be the highest
point at which the horizontally averaged $f_h$ deviates from one by a
small amount.  We will denote horizontal averages of quantities using
$\langle\cdot\rangle$, e.g., the horizontal average of $f_h$ is
$\langle f_h\rangle$.  In SG07, the height is defined to be the
highest point at which $\langle f_h\rangle$ is less than 0.985.  This
corresponds to the highest point at which the density is less than
$\rho=2.97$.  Using a lower cutoff density (or equivalently, a smaller
value of $f_h$) does not qualitatively change our results.

In Table \ref{tab:nocond} we list our simulations which have no
conductivity.  In run SG we use the parameters reported in SG07.  In
the simulations beginning with RT, we use the initial conditions
described in \S \ref{sec:numerics}.  We use the suffix LR to denote
low resolution runs with resolutions of 64x64x128 and HR to denote
high resolution runs with resolutions of 256x256x512.  RT is our
fiducial model which we will compare to our simulations with
anisotropic conduction.

In the runs labelled RT and the simulations with anisotropic conduction
presented in \S \ref{sec:cond}, the density is a function of height,
but the temperature is initially constant in the upper and lower
layers.  Thus, we will define the height to be the highest point at
which $\langle T\rangle$ differs from $T_+$ by more than 1 per cent.  This is
equivalent to the definition of height used by SG07 provided that the
pressure does not change significantly through the simulation.  In run
SG, we use the same definition of height as SG07.

\begin{figure}
\includegraphics[width=0.45\textwidth]{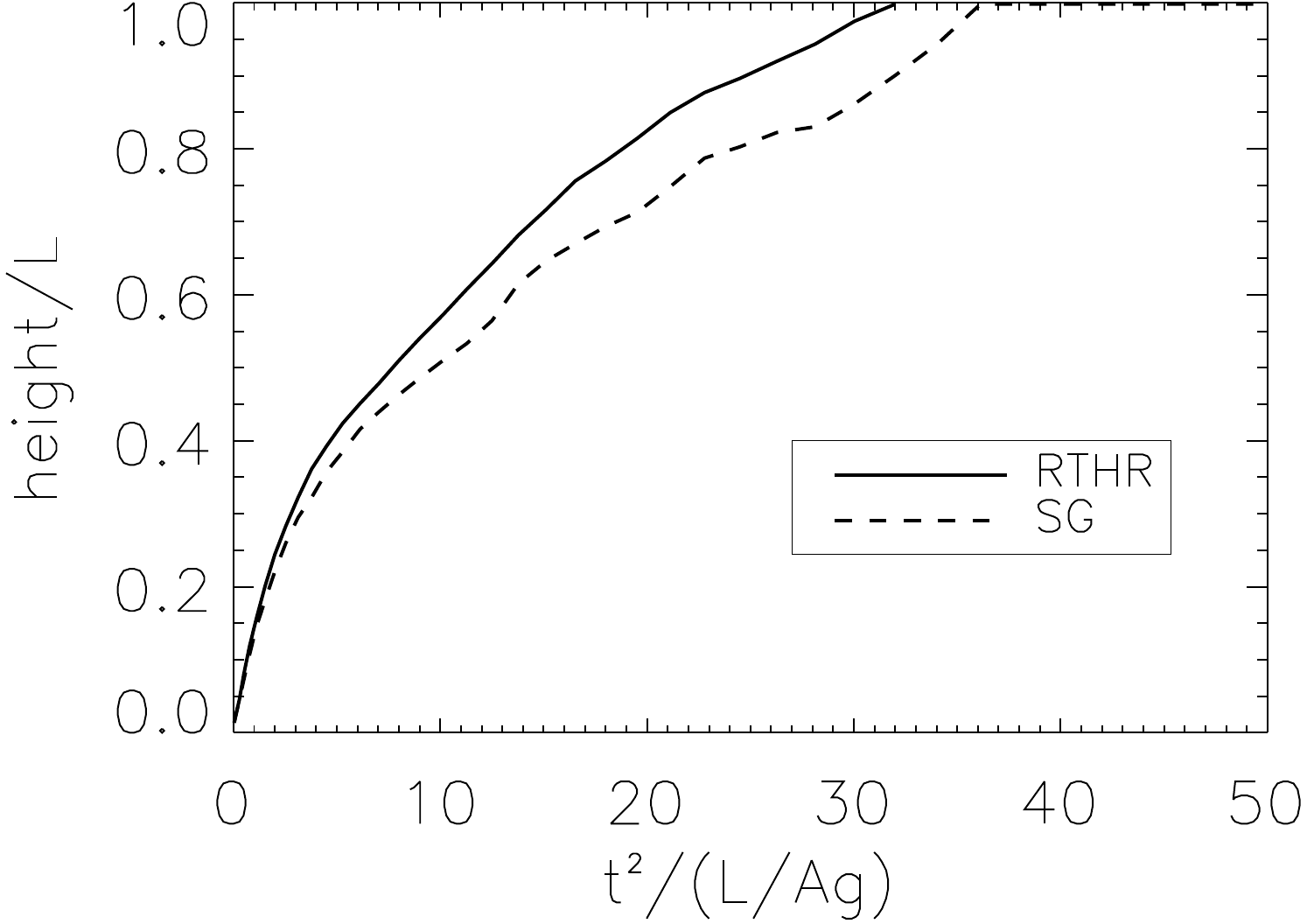}
\includegraphics[width=0.45\textwidth]{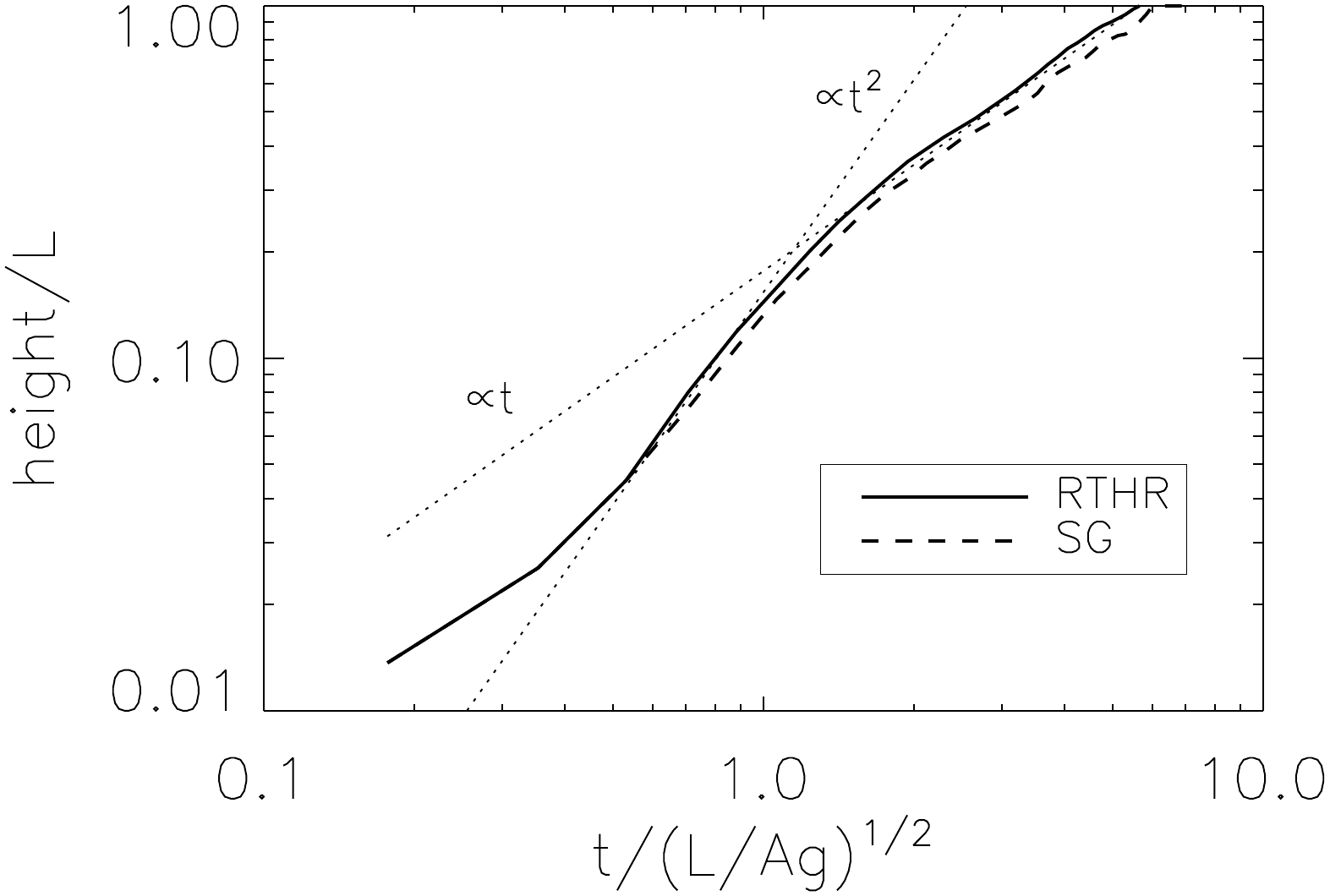}
\caption{The height of the highest bubble (defined in \S \ref{sec:RTInocond}) as a
  function of time in RTI simulations without conduction.  In the top panel
  the height is plotted against $t^2$.  In the bottom panel the height
  is plotted on a log-log plot against $t$.  We show the results from
  our high resolution simulation without conduction (solid line; RTHR in
  Table \ref{tab:nocond}) and our simulation using the parameters of
  SG07 (dashed line; SG).  The two dotted
  lines show power law relations with exponents of one and two.
  Although the height vs. $t^2$ line in the upper panel might look
  fairly linear, the bottom panel shows that the height grows linearly
  with $t$ at later times.}
\label{fig:rtheight}
\end{figure}

We plot the height for runs SG (using the same parameters as SG07) and
RTHR (our high resolution simulation without conduction) in the top
panel of Fig. \ref{fig:rtheight}.  The two simulations give similar
results.  Interestingly, they both differ somewhat from the data in
SG07 (their fig. 8).  It is unexpected that run SG has a different
height than the corresponding simulation in SG07, as we have used the
exact same parameters and a newer version of the same code as reported
in SG07.  Nevertheless, our results are qualitatively similar.  The
differences between our results and those of SG07 are presumably due
to a small difference in initial conditions or parameters of the run
that we have not been able to identify.

In the upper panel of Fig. \ref{fig:rtheight} we plot $h/L$ as a function
of $Agt^2/L$, as is often done in the literature
\citep[e.g.,][]{dim04}.  We expect to have $h=\alpha_h Agt^2$, so we
expect a straight line in this figure.  Near $t=0$, the slope is
changing rapidly, but the lines are fairly linear at later
times.  However, it is very difficult to tell from a plot of $h$
vs. $t^2$ whether or not the height is truly increasing as $t^2$.

To test whether $h$ is in fact increasing as $t^2$, we plot $h$
vs. $t$ on a log-log plot in the bottom panel of
Fig. \ref{fig:rtheight}.  The dotted lines show power laws with
exponents of one and two.  Although the two simulations are initially
roughly consistent with $h\sim t^2$, for times past
$t/\sqrt{L/Ag}\approx 1.3$, the results are more consistent with the
height increasing linearly with time.  The numerical results from SG07
are also consistent with the height increasing linearly with time.

\begin{figure}
\includegraphics[width=0.45\textwidth]{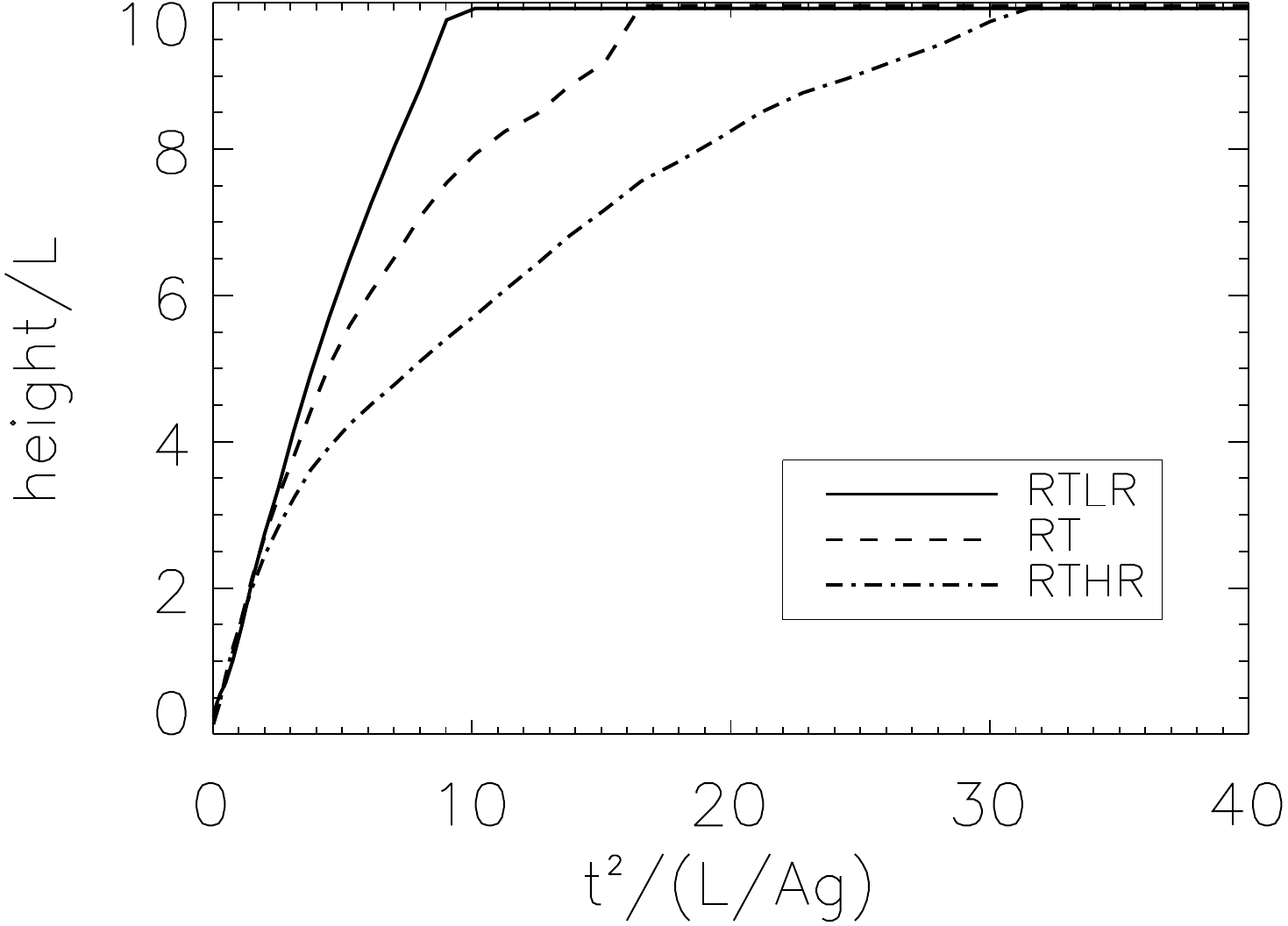}
\includegraphics[width=0.45\textwidth]{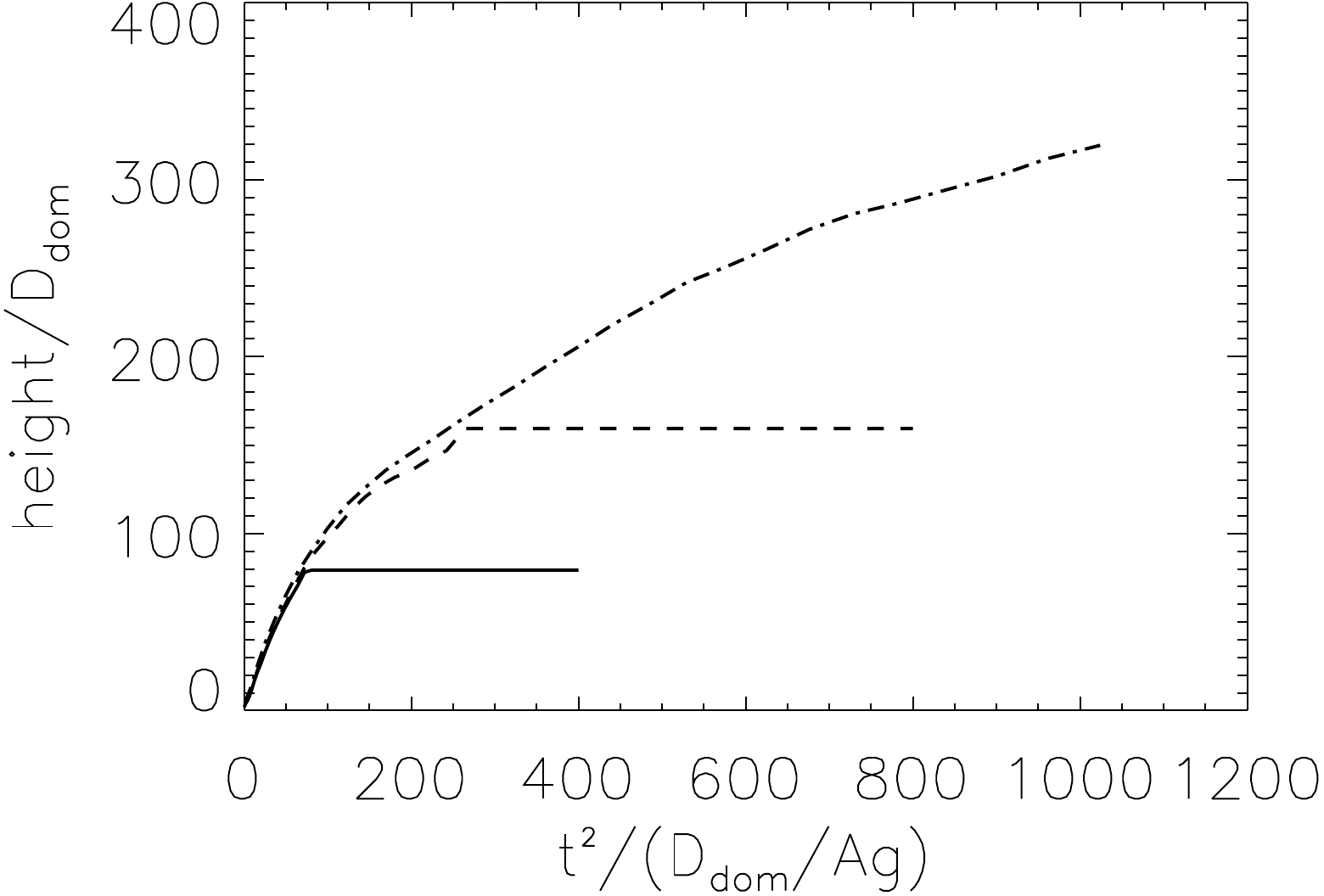}
\caption{The height of the highest bubble (defined in \S \ref{sec:RTInocond}) as a
  function of time for our RTI simulations without conduction at different
  resolutions.  In the upper panel the height and $Agt^2$ are
  normalized by the length of the box $L$.  In the lower panel, the
  height and $Agt^2$ are normalized by $D_\mathrm{dom}$, the size of
  the first dominant bubbles.  We plot our results for our low
  resolution run (solid line; RTLR in Table \ref{tab:nocond}), our
  fudicual run (dashed line; RT), and our
  high resolution run (dot-dashed line; RTHR).  We estimate $D_\mathrm{dom}$ to be $L/32$,
  or about 8 cell widths for our high resolution simulation.  We then
  take $D_\mathrm{dom}=L/16$ and $L/8$ for our fiducial simulation and
  low resolution simulation respectively.  The results line up well
  when normalizing lengths to $D_\mathrm{dom}$, but there is little
  agreement between the different resolutions when normalizing lengths
  to $L$.  This indicates that the simulations do not notice the size
  of the box, but instead remember the size of the first dominant modes.  The
  flat portion of the curves correspond to when the perturbations have
  hit the top of the box in each of the simulations.}
\label{fig:rtres}
\end{figure}

In Fig. \ref{fig:rtres} we plot, for three different resolutions, the
height normalized to $L$ as a function of $Agt^2$ normalized to $L$.
There is no significant agreement between the simulations at different
resolutions.  However, in the bottom panel of Fig. \ref{fig:rtres}, we
show the height normalized to $D_\mathrm{dom}$ as a function
of $Agt^2$ normalized to $D_\mathrm{dom}$, where $D_\mathrm{dom}$ is
the size of the first dominant bubbles and spikes in each simulation.
There is much closer agreement between the runs at different
resolutions using this normalization.  This effect was previously noted
by, e.g., \citet{dim04}.

The statement that $h$ should increase as $t^2$ is often motivated by
dimensional analysis: if the only dimensional parameter in the problem
is $g$, then we must have that $h=Cgt^2$ for some dimensionless
coefficient $C$ (which is related to the Atwood number).  However, our simulations
show there is another dimensional number in the problem:
$D_\mathrm{dom}$.  When we increase the resolution of our simulations
by a factor of two in each direction, $D_\mathrm{dom}$ decreases by a
factor of about two.  The bottom panel of Fig. \ref{fig:rtres} shows
that the relevant length scale for the non-linear evolution is
$D_\mathrm{dom}$, and indicates that the simulations remember the size
of the initial dominant modes very late into the non-linear regime.

\begin{figure}
\includegraphics[width=0.45\textwidth]{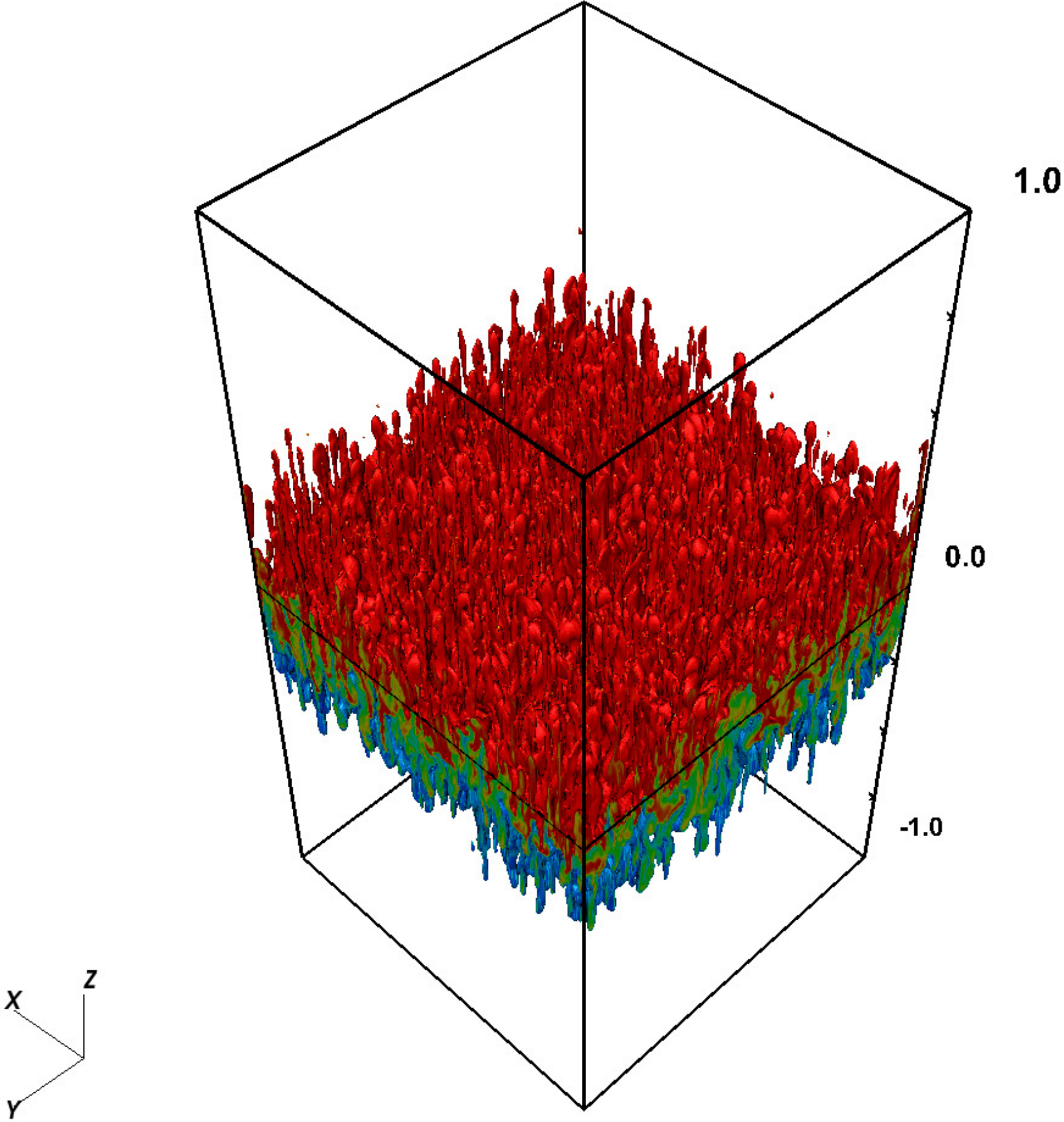}
\includegraphics[width=0.45\textwidth]{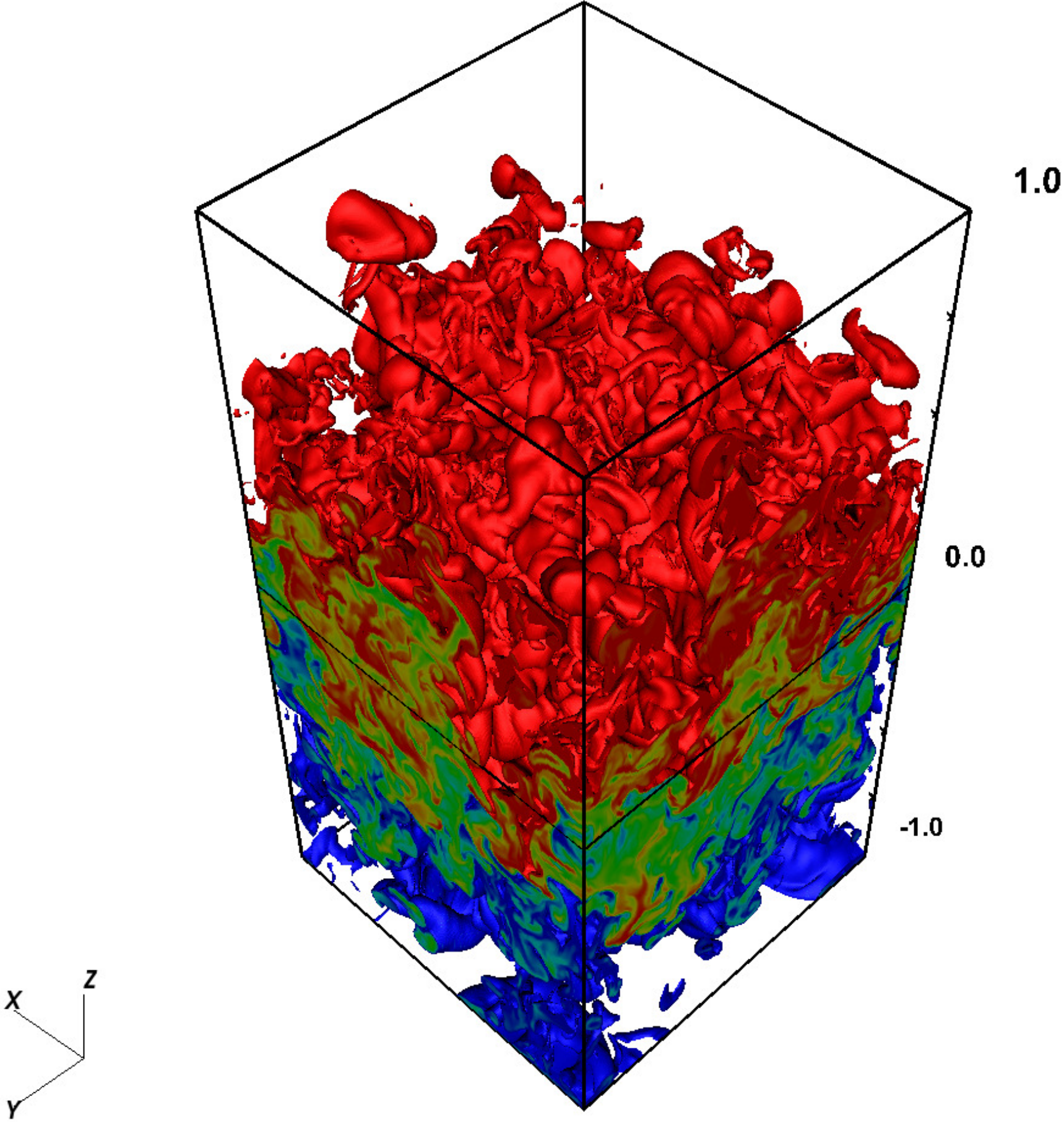}
\caption{Isosurfaces of temperature at $T=95$ and $T=38$ for the high
  resolution simulation without any explicit conductivity (RTHR in
Table \ref{tab:nocond}).  The upper (lower) layer is initialized with
a temperature of 33 (100).  The upper panel is at time
$t/\sqrt{L/Ag}\approx 1.8$, and the lower panel is at time
$t/\sqrt{L/Ag}\approx 4.2$.  Temperature contours are also shown on
the faces of the simulation domain.  The height of the highest bubble
is increasing linearly at both times shown (see Fig. \ref{fig:rtheight}),
consistent with a bubble of fixed size rising at its terminal
velocity.  However, the size of the bubbles is increasing, which would
yield larger terminal velocities at later times.}
\label{fig:rtcontours}
\end{figure}

With another dimensional number in the problem, $D_\mathrm{dom}$, we
can make a new \textit{dimensionless} parameter,
$D_\mathrm{dom}/(gt^2)$.  With this additional dimensionless
parameter, the height is no longer required to increase as $gt^2$.
Instead, for instance, the height could increase as
$t\sqrt{D_\mathrm{dom}g}$.  Physically, this would correspond to a
bubble of fixed size rising at its terminal velocity.  This functional
form fits our numerical results, although the size of the
bubbles in our simulation increase as a function of time (see
Fig. \ref{fig:rtcontours}).

Although we do not have a satisfactory explanation for why the height
increases linearly with time, it is clear that our results are not
consistent with the commonly repeated result that the height increases
quadratically with time.  We have also shown that the results of SG07
are also not consistent with the height increasing quadratically with
time (although SG07 did not make this point explicitly).  This shows
that care must be taken when determining the functional form of
$h(t)$.
 
With this in mind, we will now include anisotropic conductivity in our
calculations.  We will be making frequent comparison with our fiducial
RTI simulation without explicit conduction (RT in
Table \ref{tab:nocond}), as well as with the results shown in
Figures \ref{fig:rtheight}--\ref{fig:rtcontours}.

\section{Results with Anisotropic Conduction}\label{sec:cond}

Anisotropic conduction enables diffusive spreading of the initial
temperature discontinuity.  Even horizontal fields induce diffusive
spreading -- the small $v_z$ from the initial perturbation produces a
small vertical field component.  This produces a small amount of
perpendicular temperature diffusion when the interface between the two
constant temperature layers is within a cell.  To distinguish mixing
of temperature from mixing of material on either side of the
interface, we will introduce two new definitions of height.  We will
call the definition used in
\S \ref{sec:RTInocond} the height of the temperature mixing layer,
or $h_T$.  As before, it is defined to be the highest point at which
$\langle T\rangle$ differs from $T_+$ by more than 1 per cent.  We will also
sometimes calculate the depth of the temperature mixing layer, or
$d_T$.  Similarly, $d_T$ is the lowest point at which $\langle
T\rangle$ differs from $T_-$ by more than 1 per cent.

In addition, we initialize the simulations with two passive scalar
fields which we will call $C_L$ and $C_U$.  The simulations start with
$C_L=1$, $C_U=0$ for $z<0$ and with $C_L=0$, $C_U=1$ for $z>0$.  We
define the two heights of the composition mixing layer, $h_{C_L}$ and
$h_{C_U}$, to be the highest points for which $\langle C_L\rangle$ or
$\langle C_U\rangle$ differ from their initial value by more than
0.015.  That is, $h_{C_L}$ is the highest point at which $\langle
C_L\rangle > 0.015$ and $h_{C_U}$ is the highest point at which
$\langle C_U\rangle < 0.985$.  When there is no temperature diffusion,
$h_T=h_{C_L}=h_{C_U}$ provided that $\rho_+=3$ and $\rho_-=1$, so the
height of the temperature mixing layer, the two heights of the
composition mixing layer, and the height of the highest RTI bubble all
coincide.

In Table \ref{tab:cond} we list the simulations with conduction that
we will discuss in this paper.  We studied interfaces with RTI-stable
or RTI-unstable jumps in temperature, as well as magnetic fields which
are initially horizontal, vertical, or at a 45 degree angle to the
horizontal (see Fig. \ref{fig:chart}).  The names of simulations with
initially horizontal magnetic fields start with `H' in
Table \ref{tab:cond}, whereas the names of simulations with initially
vertical magnetic fields start with `V.'  The two simulations with an
initial magnetic field at a 45 degree angle to the horizontal have
names starting with `A.'  Simulations which are RTI stable have an `S'
as the second character in their name.  We vary the thermal
diffusivity, which is given by $\chi=10^{-n}$, where $n$ is the number
in the name of the simulation (in simulations with initially skewed
magnetic fields, $\chi=2\times 10^{-n}$).  A diffusivity of $10^{-n}$
corresponds to $\omega_c/\omega_{\mathrm{RTI}}=9\times 10^{-n}$, where
$\omega_c$ is conduction frequency over the simulation domain, and
$\omega_{\mathrm{RTI}}$ is the RTI growth rate for modes the scale of
the simulation domain.  Some of our simulations employ isotropic
instead of anisotropic conduction -- these have the letter `I' in their
names.  Most of our simulations are at a resolution of 128x128x256
(the simulations with RTI-stable temperature jumps and vertical
magnetic field are run for longer, so they use a resolution of
64x64x128).  However, to test resolution effects, we ran some
simulations at a resolution of 256x256x512 (we ran the simulations
with RTI-stable temperature jumps with vertical magnetic fields at a
resolution of 128x128x256) -- these high resolution runs are denoted by
`HR.'

\begin{table}
\caption{Parameters for simulations with conductivity.}
\label{tab:cond}
\begin{center}
\begin{tabular}{@{}ccccc}
Name & Resolution & $\omega_c/\omega_{\mathrm{RTI}}$ & $\boldsymbol{B}$ dir. & RTI stability \\
\hline
H0 & 128x128x256 & 9 & hor. & unstable \\
H1 & 128x128x256 & 0.9 & hor. & unstable \\
H2 & 128x128x256 & 0.09 & hor. & unstable \\
H2HR & 256x256x512 & 0.09 & hor. & unstable \\
HS0 & 128x128x256 & 9 & hor. & stable \\
HS1 & 128x128x256 & 0.9 & hor. & stable \\
HS2 & 128x128x256 & 0.09 & hor. & stable \\
HS2HR & 256x256x512 & 0.09 & hor. & stable \\
V1 & 128x128x256 & 0.9 & ver. & unstable \\
V2 & 128x128x256 & 0.09 & ver. & unstable \\
V3 & 128x128x256 & 0.009 & ver. & unstable \\
V3HR & 256x256x512 & 0.009 & ver. & unstable \\
VS2 & 64x64x128 & 0.09 & ver. & stable \\
VS3 & 64x64x128 & 0.009 & ver. & stable \\
VS3HR & 128x128x256 & 0.009 & ver. & stable \\
VS4 & 64x64x128 & 0.0009 & ver. & stable \\
VS1I & 128x128x256 & 0.9 (i) & ver. & stable \\
VS2I & 128x128x256 & 0.09 (i) & ver. & stable \\
VS3I & 64x64x128 & 0.009 (i) & ver. & stable \\
VS3IHR & 128x128x256 & 0.009 (i) & ver. & stable \\
VS4I & 128x128x256 & 0.0009 (i) & ver. & stable \\
A3 & 128x128x256 & 0.018 & $45^\circ$ & unstable \\
AS2 & 128x128x256 & 0.18 & $45^\circ$ & stable \\
\hline
\end{tabular}
\end{center}
\medskip
The simulations are run as described in \S \ref{sec:numerics}.  The
conductivity is anisotropic except for the simulations labelled (i).
$\omega_c$ is the conduction time-scale across the simulation domain
and $\omega_{\mathrm{RTI}}$ is the RTI growth rate for modes with
wavelengths comparable to the size of the simulation domain.  If the
simulation is listed as RTI-unstable, then $\rho_+=3$ and $\rho_-=1$.
If the simulation is listed as RTI-stable then $\rho_+=1$ and
$\rho_-=3$.
\end{table}

\subsection{Horizontal magnetic field}\label{sec:horB}

Systems with horizontal magnetic fields and anisotropic conduction are
sometimes susceptible to the MTI.  The MTI and RTI both occur when
$\rho_+>\rho_-$.  We will start by discussing the case where
$\rho_+<\rho_-$.  There is neither MTI nor RTI in this case, so the
dynamics are dominated by diffusion perpendicular to the magnetic
field lines due to a combination of small perturbations to the field
orientation and numerical mixing across the interface.  If a
simulation is run with no initial perturbation, we find that the code
holds the equilibrium exactly.

\subsubsection{RTI-stable and MTI-stable ($\rho_+<\rho_-$)}\label{sec:horstable}

\begin{figure}
\includegraphics[width=0.45\textwidth]{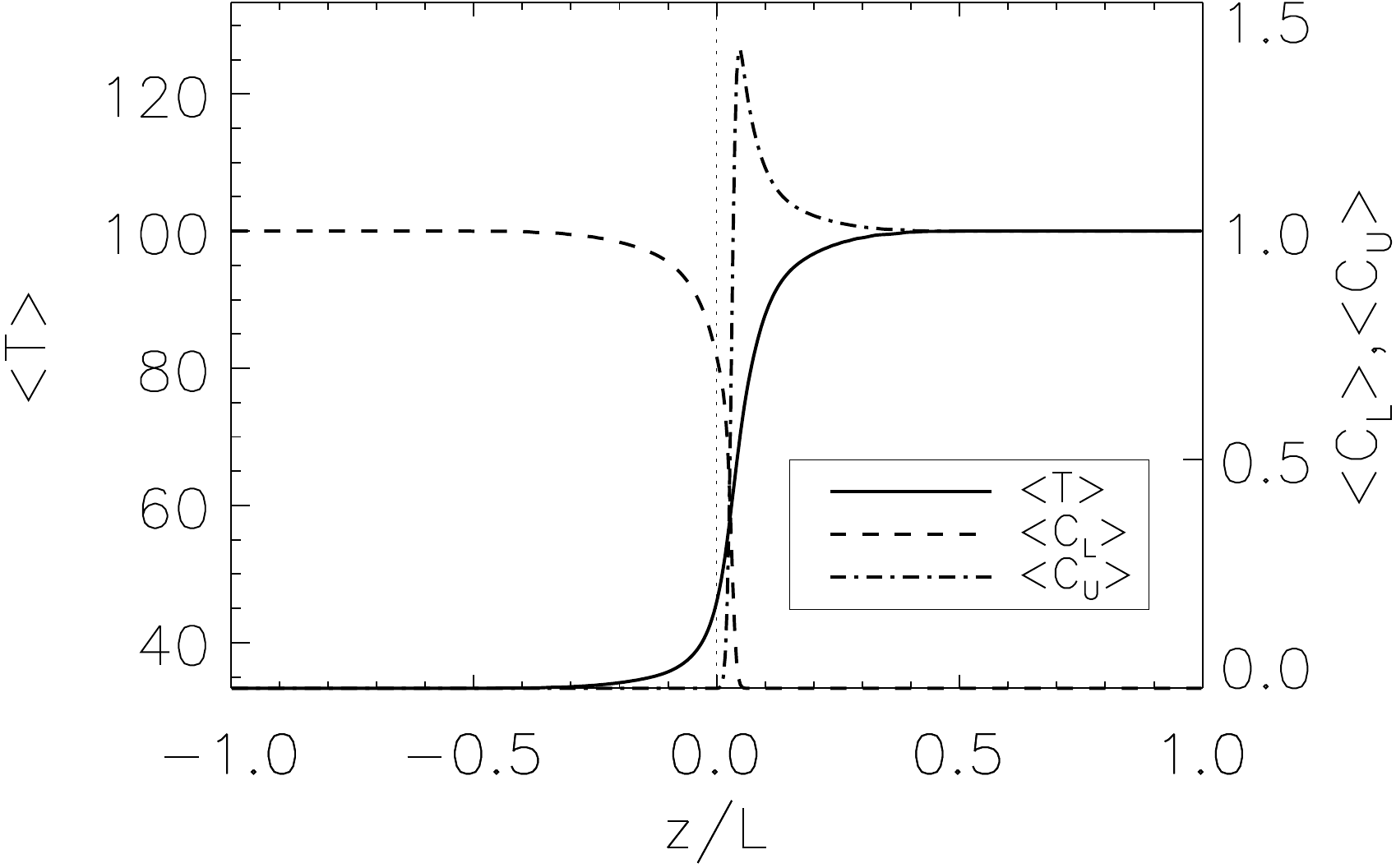}
\caption{Profiles of $\langle T\rangle$, $\langle C_L\rangle$ and $\langle C_U\rangle$
   for our high resolution RTI-stable simulation with
  $\omega_c/\omega_\mathrm{RTI}=0.09$ (HS2HR in Table \ref{tab:cond})
  at $t/\sqrt{L/Ag}\approx 3.5$.  $C_L$ ($C_U$) is a passive scalar
  initialized to one (zero) for $z<0$ and zero (one) for $z>0$ which
  track the mixing of density, rather than the mixing of temperature
  (which is affected by heat diffusion).  The vertical dotted line is
  at $z=0$.  The interface (about where $\langle C_L\rangle=\langle
  C_U\rangle$) has moved upwards from $z=0$, compressing $C_U$ above
  the interface and spreading out $C_L$ below the interface.  This
  upwards motion is due to a pressure force caused by the small
  vertical heat diffusion in the simulation.}
\label{fig:rthsheights}
\end{figure}

Because there is neither MTI nor RTI the distinction between the
height of the temperature mixing layer ($h_T$) and the composition
mixing layer ($h_{C_L}$ and $h_{C_U}$) is very important.  In
Fig. \ref{fig:rthsheights} we plot $\langle T\rangle$, $\langle
C_L\rangle$, and $\langle C_U\rangle$ as a function of $z$ at
$t/\sqrt{L/Ag}\approx 3.5$ for our high resolution run with
$\omega_c/\omega_\mathrm{RTI}=0.09$ (HS2HR in Table \ref{tab:cond}).
Several effects can be seen in this figure.  First, the vertical
position of the interface (about where $\langle C_L\rangle$ and
$\langle C_U\rangle$ intersect) is above $z=0$.  Because the interface
is moved upwards, the upper layer is compressed, causing $C_U$ to build
up above the interface.  Note that there is very little movement of
material across the interface -- the gradient of $\langle C_U\rangle$
remains large throughout the simulation.  Similarly, $\langle
C_L\rangle$ does not spread upwards across the interface.  However,
because the interface is moving upwards, $C_L$ is spreading out below
the interface.  In fact, the depth associated with $\langle
C_L\rangle$ is equal to the depth associated with $\langle T\rangle$.
Also notice that our metrics for the height of the composition mixing
layer are sensitive to both the width of the composition mixing layer
and the vertical position of the centre of the mixing layer.  Thus,
the size of the mixing layer is actually being overestimated by
$h_{C_L}$ and $h_{C_U}$, despite remaining small throughout the
simulations.  The results depicted in Fig. \ref{fig:rthsheights} are
similar for other times -- at earlier times, the interface is closer
to $z=0$, whereas at later times, the interface is at a higher
vertical position.

The origin of the upwards motion of the interface can be understood as
follows.  Because of the small initial perturbation to $v_z$ and
numerical mixing across the interface, there is some vertical
conduction of heat.  The perpendicular conduction is proportional to
the amplitude of the initial perturbation which adds a small vertical
component to the magnetic field.  This vertical heat conduction will
produce a vertical pressure force in the following manner.  Consider
the dynamics at the beginning of the simulation and very close to the
interface at $z=0$.  Initially the temperature above the interface is
higher than the temperature below the interface, while the pressure is
about equal above and below the interface.  Dynamically nothing is
happening because the interface is stable to dynamical perturbations.
However, there is some vertical heat conduction, so on a temperature
diffusion time-scale, heat is transferred from above the interface to
below the interface.  Since the temperature above the interface drops,
the pressure above the interface also drops.  Similarly, the pressure
below the interface increases.  This pressure difference pushes the
interface upwards.

\begin{figure}
\includegraphics[width=0.45\textwidth]{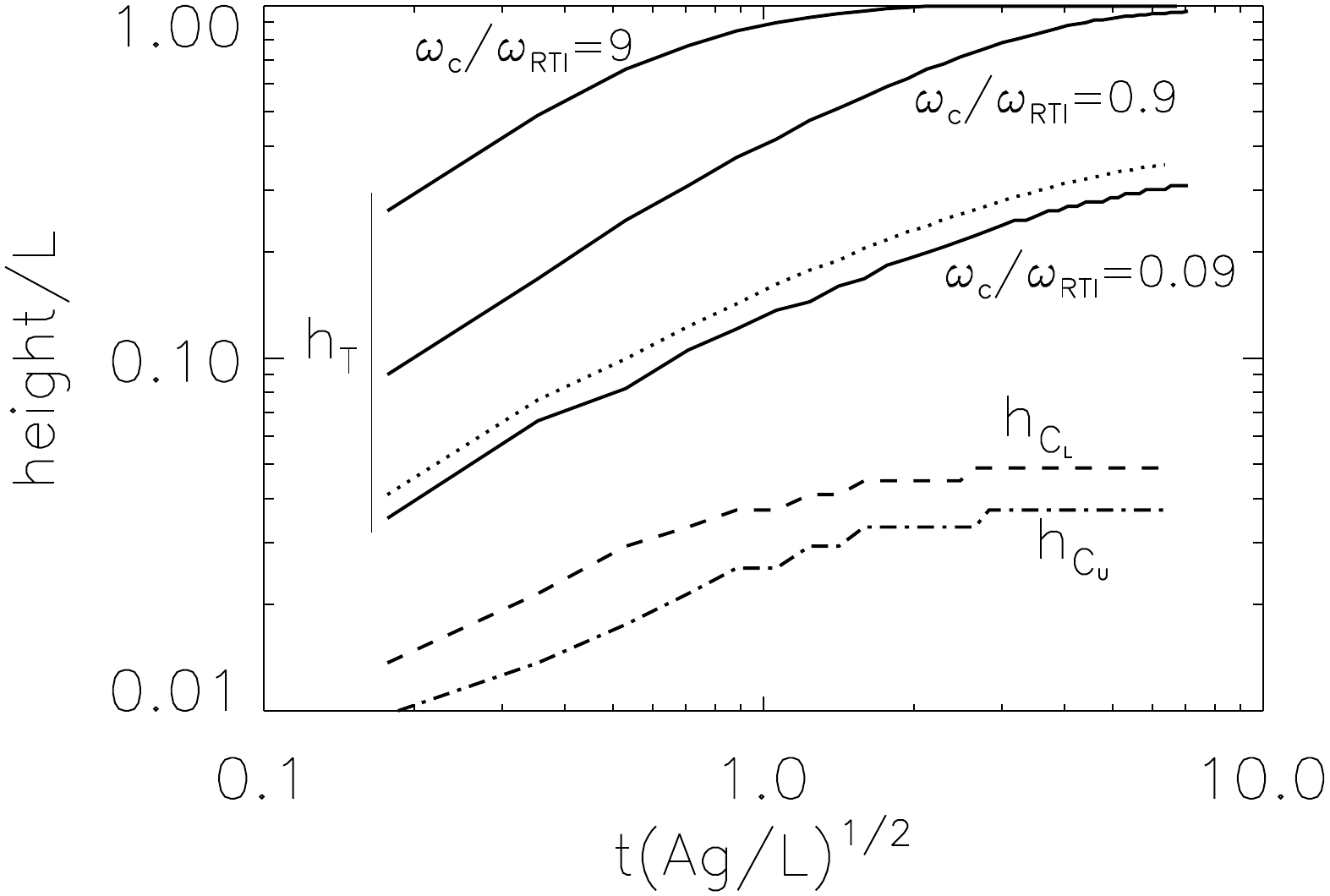}
\caption{Heights (defined in
  \S \ref{sec:cond}) as a function of time for various simulations
  with horizontal magnetic fields and an RTI-stable temperature jump.
  The solid and dotted lines are plots of $h_T$, the height of the
  temperature mixing layer.  The highest solid line correspond to the
  run with $\omega_c/\omega_\mathrm{RTI}=9$ (HS0 in
  Table \ref{tab:cond}), the middle solid line correspond to the run
  with with $\omega_c/\omega_\mathrm{RTI}=0.9$ (HS1), and the lowest
  solid line and dotted line correspond to the two runs with
  $\omega_c/\omega_\mathrm{RTI}=0.09$ (solid line is at the fiducial
  resolution, HS2; dotted line is the high resolution run, HS2HR).
  The dashed and dot-dashed lines show $h_{C_L}$ and $h_{C_U}$
  respectively for simulation HS2HR; these are two metrics for the
  height of the composition mixing layer.  If there was no
  perpendicular temperature diffusion, all heights would stay at zero
  for the entire simulation.  Although little material diffuses across
  the interface, $h_{C_L}$ and $h_{C_U}$ increase slowly due to a
  pressure force which raises the interface.}
\label{fig:rthstable}
\end{figure}

In Fig. \ref{fig:rthstable} we plot the height as a function of time
for several different values of $\omega_c/\omega_\mathrm{RTI}$.
$h_{C_L}$ and $h_{C_U}$ are only plotted for the high resolution
simulation (HS2HR in Table \ref{tab:cond}), but these heights are
similar for the other simulations.  As the conductivity increases, all
the height metrics increase, although $h_{C_L}$ and $h_{C_U}$ remain
about an order of magnitude smaller than $h_T$.  Note that $h_T$ is
slightly larger at higher resolutions.  This might be because the
upwards pressure force due to conduction is larger for smaller cell
widths, as the initial gradient is sharper.  We have defined $h_{C_L}$
and $h_{C_U}$ to be the points at which $\langle C_L\rangle$ and
$\langle C_U\rangle$ differ by 1.5\%, which corresponds to a 1 per cent
temperature variation if $\rho_+=3$.  However, in these simulations
$\rho_+=1$, so if $T, C_L,$ and $C_U$ were all mixed by the same
amount, we would expect $h_{C_L}=h_{C_U}$ to be larger than $h_T$,
instead of much smaller than $h_T$ as shown in
Fig. \ref{fig:rthstable}.  This highlights the fact that the dominant
effect in these RTI-stable simulations is simple temperature diffusion
which increases $h_T$ much more than $h_{C_L}$ or $h_{C_U}$.

\begin{figure}
\includegraphics[width=0.45\textwidth]{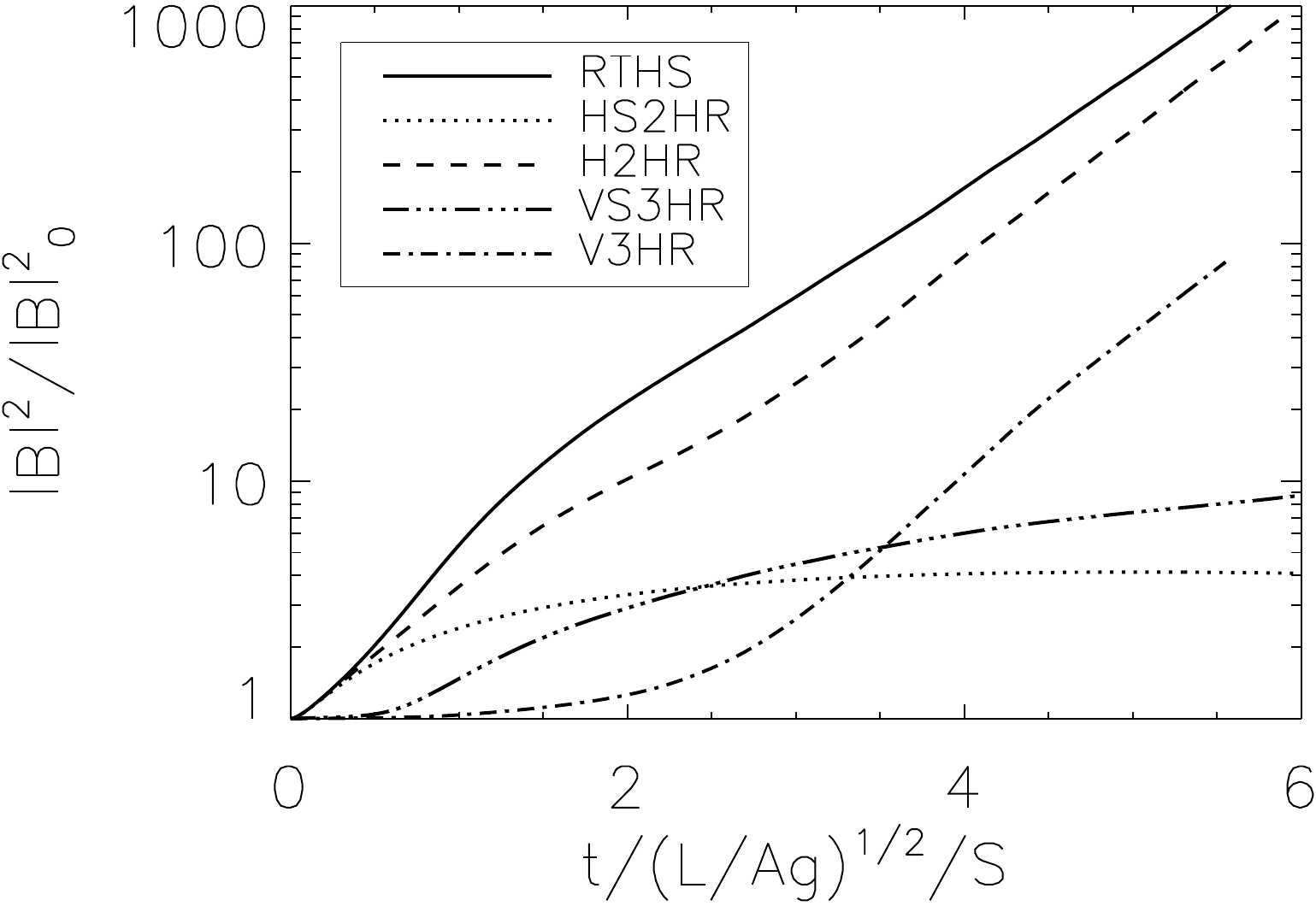}
\caption{The magnetic energy, $|\boldsymbol{B}|^2/8\pi$, normalized to
  the initial magnetic energy as a function of time.  We show
  results from our high resolution simulation without any
  explicit conductivity (solid line; RTHR in Table \ref{tab:nocond});
  our high resolution simulations with initially horizontal magnetic
  fields with $\omega_c/\omega_\mathrm{RTI}=0.09$, one which is
  RTI-stable (dotted line; HS2HR in Table \ref{tab:cond}), the other
  which is RTI-unstable (dashed line; H2HR); and our high resolution
  simulations with initially vertical magnetic fields with
  $\omega_c/\omega_\mathrm{RTI}=0.009$, one which is RTI-stable
  (triple-dot-dashed line; VS2HR), and one which is RTI-unstable
  (dot-dashed line; V2HR).  The HBI-unstable simulation, VS2HR, is run
  for much longer than the other simulations; for convenience, we
  rescale the time variable using a scaling factor $S$ which is equal
  to 12 for simulation VS2HR and 1 for all other simulations.  The
  RTI-unstable simulations exhibit exponentially growing magnetic
  energies, whereas the RTI-stable simulations show linearly
  increasing magnetic energies.}
\label{fig:ME}
\end{figure}

We also tracked the amplification of magnetic energy in the
simulations.  The results are shown in Fig. \ref{fig:ME}.  In this
figure, we plot the growth of magnetic energy for all our high
resolution simulations.  For comparison, we include the high
resolution simulation without any explicit conductivity (solid line).
In the RTI-unstable simulation without any explicit conductivity, the
magnetic energy grows exponentially for most of the simulation.  This
is because the magnetic field remains dynamically unimportant for the
entire simulation.  In our high resolution RTI-stable simulation with
a horizontal initial magnetic field, there is some initial linear
growth in the magnetic energy which saturates by $t/\sqrt{L/Ag}\approx
2$; the magnetic energy increases by a factor of four.  This is
likely due to the initial velocity perturbation.  Unsurprisingly, the
growth is significantly smaller than in the RTI-unstable simulation,
although the two cases have the same initial magnetic field.

These results highlight the essential role of perpendicular thermal
conduction in our RTI-stable simulations with horizontal magnetic
fields.  Without the perpendicular conduction, there would be no
vertical temperature diffusion, so $h_T$ would stay close to zero for
the entire simulation.  The interface would also not rise so $h_{C_L}$
and $h_{C_U}$ would also stay close to zero.  In any astrophysical
context, however, there will always be some diffusion across the
interface.  Temperature diffusion across the interface could occur due
to some small perpendicular diffusion, or because the magnetic field
is not completely parallel to the interface.  In these cases the
system will evolve as shown in
Figures \ref{fig:rthsheights} \& \ref{fig:rthstable}, although the
time scale for diffusion will depend on the source and strength of
diffusion across the interface.

\subsubsection{RTI-unstable and MTI-unstable ($\rho_+>\rho_-$)}\label{sec:horunstable}

\begin{figure}
\includegraphics[width=0.45\textwidth]{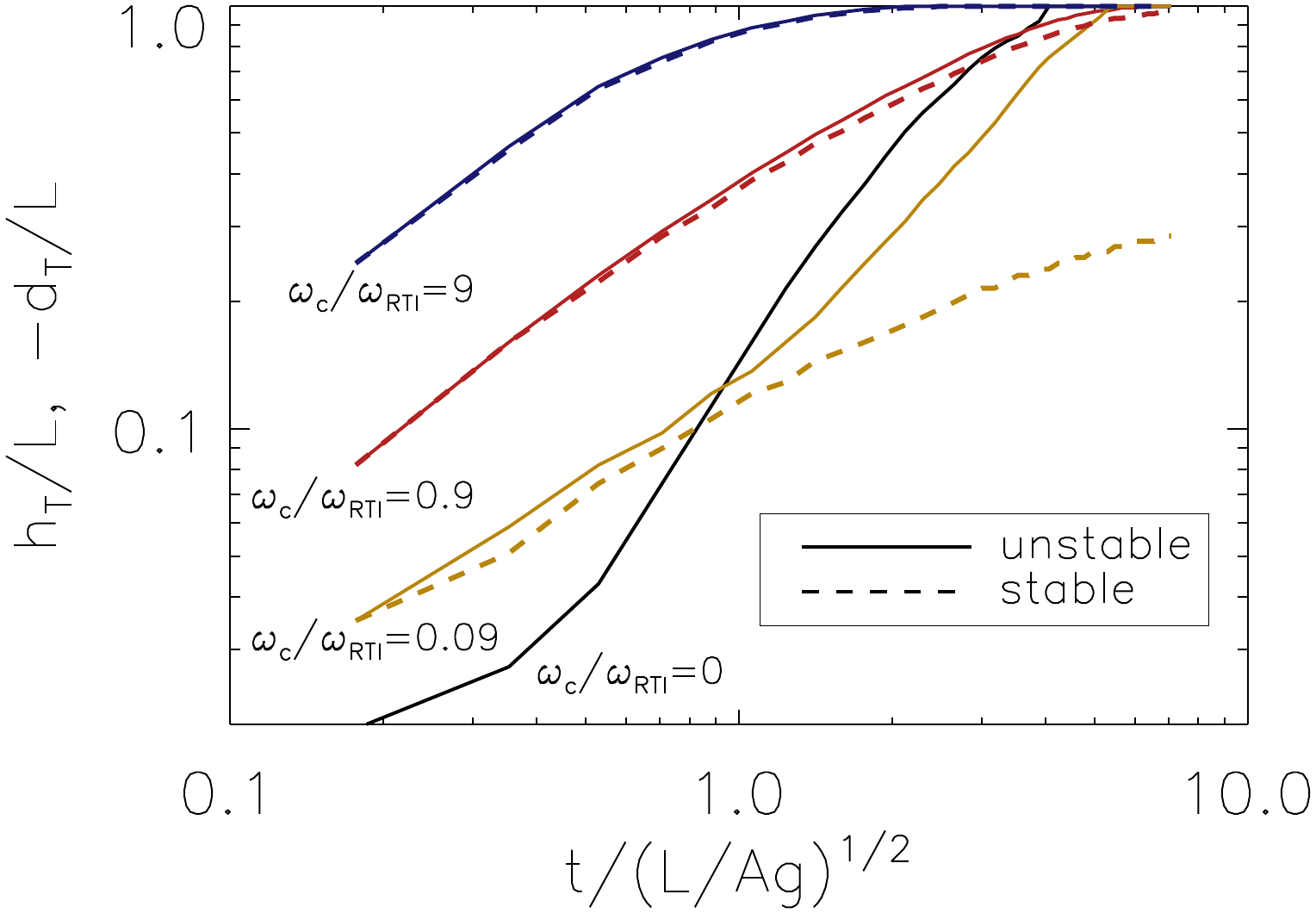}
\caption{Heights and depths of the temperature mixing layer (defined in
  \S \ref{sec:cond}) as a function of time for various
  $\omega_c/\omega_\mathrm{RTI}$ for simulations with horizontal
  initial magnetic fields.  The solid lines show $h_T$ for our
  simulations which are RTI-unstable, whereas the dotted lines show
  $-d_T$ for our simulations which are RTI-\textit{stable}.  The top
  pair of blue lines, which are almost overlapping are for
  $\omega_c/\omega_\mathrm{RTI}=9$ (H0 and HS0 in
  Table \ref{tab:cond}).  Below these, the red pair of lines, which
  are also almost overlapping are for
  $\omega_c/\omega_\mathrm{RTI}=0.9$ (H1 and HS1).  The lowest pair of
  lines, which are in yellow, correspond to
  $\omega_c/\omega_\mathrm{RTI}=0.09$ (H2 and HS2).
  The bottom solid line shows $h_T$ for our fiducial simulation without
  explicit conductivity (RT in Table \ref{tab:nocond}).  The height for
  the RTI-unstable runs is about equal to the depth for the RTI-stable
  runs when conduction dominates over the RTI, which is for most of
  the simulation when $\omega_c/\omega_\mathrm{RTI}=9$ and $0.9$.
  However, for $\omega_c/\omega_\mathrm{RTI}=0.09$, the RTI becomes
  important around $t/\sqrt{L/Ag}\approx 1-2$, and the height in the
  RTI-unstable simulation increases in a similar way to the height in
  the simulation without conductivity.  The results are qualitatively similar at
  higher resolution.}
\label{fig:horunstable}
\end{figure}

When the density above the interface is larger than the density below
it, the contact discontinuity is RTI-unstable.  We plot $h_T$ for the
RTI-unstable simulations for several values of
$\omega_c/\omega_\mathrm{RTI}$ as solid lines in
Fig. \ref{fig:horunstable}.  We also plot the depth of the temperature
mixing layer, $d_T$, for the RTI-\textit{stable} simulations for
several values of $\omega_c/\omega_\mathrm{RTI}$ as dashed lines.
Using these two sets of lines, we can determine when the increase in
height is due to vertical heat conduction and when it is due to RTI
non-linear mixing.  Both the depth of the temperature mixing layer in
the RTI-stable case and the height of the temperature mixing layer in
the RTI-unstable case describe how temperature penetrates into
the layer with $T=100/3$.  If vertical heat conduction is more
important than the RTI, then we would expect $h_T$ for the
RTI-unstable case to equal $-d_T$ for the RTI-stable case.  Indeed, we
see that $h_T$ is very close to $-d_T$ for the entire simulation when
$\omega_c/\omega_\mathrm{RTI}=9$ and $0.9$.  This indicates that
vertical heat conduction dominates the RTI for these diffusivities.
However, when $\omega_c/\omega_\mathrm{RTI}=0.09$, the vertical heat
conduction only dominates until $t/\sqrt{L/Ag}\approx 1-2$.  After this
time, the RTI becomes more important, and the height increases as in
our fiducial simulation with no explicit conduction.

In general, vertical heat conduction is always faster than the RTI
early in the simulation because the conduction time over the length
scale associated with the discontinuity is nearly zero.  Thus, the
temperature is essentially unaffected by the RTI at early times.
However, the RTI does continue to grow, even in the
diffusion-dominated regime early in the simulations.  The RTI bubbles
grow faster than heat diffuses, so the RTI always wins at late times
provided the simulation domain is large enough.  The transition to the
RTI-dominated regime occurs when the height of the RTI bubbles reaches
the height of the temperature mixing layer due to conduction alone.
If the RTI was occurring independently of the conduction, this would
occur when $h_T$ in our fiducial simulation equals $-d_T$ for the
RTI-stable case.  This is broadly in agreement with
Fig. \ref{fig:horunstable}.

However, the height after the transition to the RTI-dominated regime is
not equal to the height in our fiducial simulation which has no
explicit conductivity.  The simulations with anisotropic conduction
are more dissipative so the height increases more slowly than when
there is no explicit conductivity.  This is the principal effect of
anisotropic conduction for the horizontal field, RTI-unstable case, in
the RTI-dominated regime.  The increase in dissipation is due to
magnetic field lines penetrating the bubbles and diffusing their
temperature into the surrounding medium.  This decreases the buoyancy
of the bubbles and slows their rise.

\begin{figure}
\includegraphics[width=0.45\textwidth]{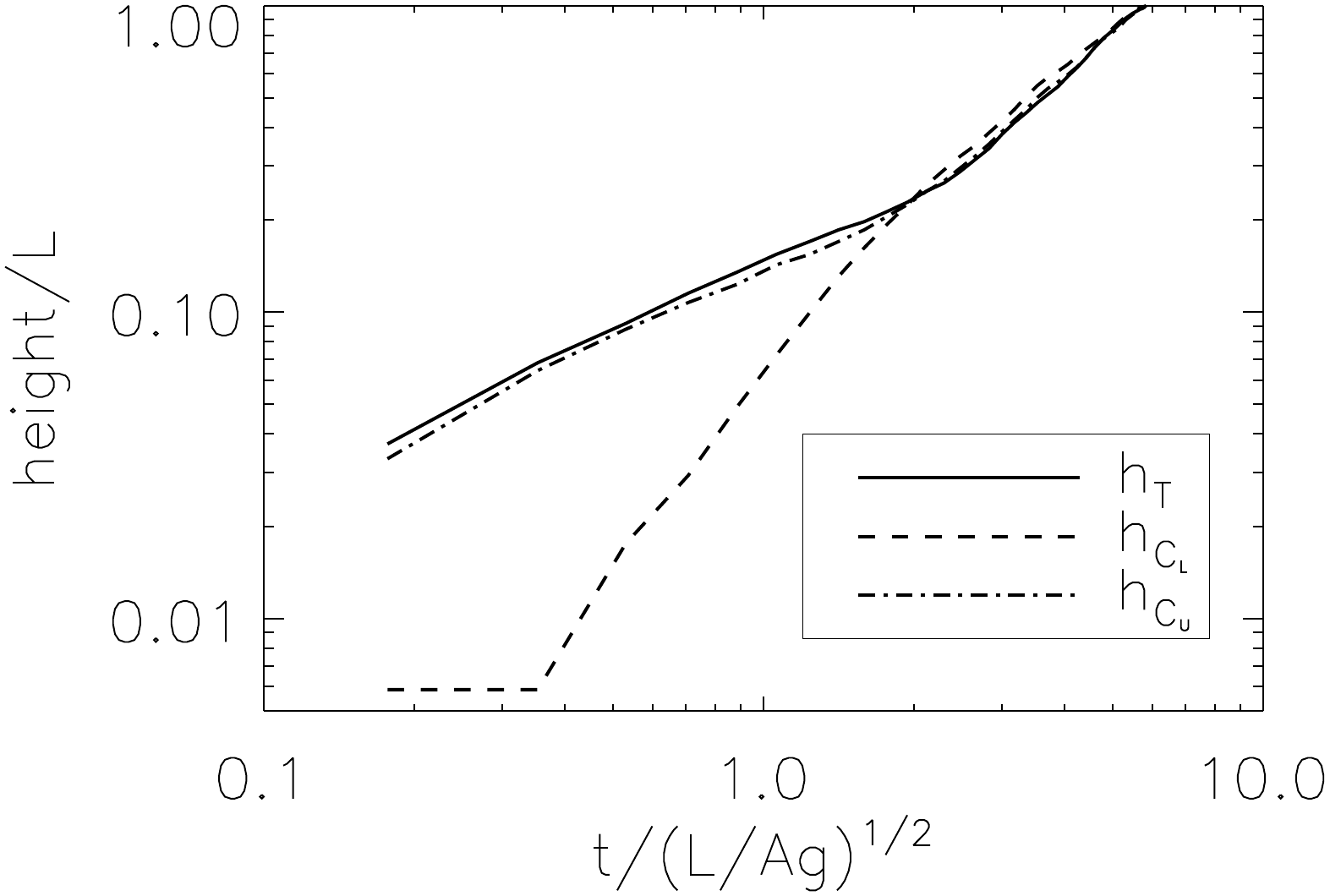}
\caption{Heights (defined in \S \ref{sec:cond}) as a function of time for our high resolution RTI-unstable
  run with an initially horizontal magnetic field, and with
  $\omega_c/\omega_\mathrm{RTI}=0.09$ (HS2HR in Table \ref{tab:cond}).
  We plot the height of the temperature mixing layer, $h_T$ (solid
  line), as well as the two heights of the composition mixing layer,
  $h_{C_L}$ and $h_{C_U}$ (dashed and dot-dashed lines respectively).
  When conduction is more important than the RTI, $h_{C_L}$ stays
  small, although $h_{C_U}$ is about equal to $h_T$.  This is because
  the profiles of $\langle T\rangle$, $\langle C_L\rangle$, and
  $\langle C_U\rangle$ look similar to those in
  Fig. \ref{fig:rthsheights}, except with $z\rightarrow -z$ and
  $C_L\leftrightarrow C_U$.  When the RTI becomes important at around
  $t/\sqrt{L/Ag}\approx 1-2$, everything becomes well mixed, and all
  three height metrics are about equal.}
\label{fig:horcomp}
\end{figure}

In the RTI-stable case ($\rho_+<\rho_-$), the height of the
composition mixing layer and the height of the temperature mixing
layer were fairly decoupled (see Fig. \ref{fig:rthstable}).  We plot our three height metrics in
Fig. \ref{fig:horcomp} for our high resolution RTI-unstable run (RT2HR
in Table \ref{tab:cond}).  We see that initially $h_{C_U}$ is about
equal to $h_T$, whereas $h_{C_L}$ is much smaller.  At later times,
once the RTI becomes important, all three height metrics become about
equal.

At early times, when conduction is more important than the RTI, the
evolution is very similar to the RTI-stable case.  However, because
the high temperature layer is on the bottom rather on top, the heights
in the RTI-unstable case match the depths in the RTI-stable case, and
$C_L$ and $C_U$ are switched.  The profiles of $\langle T\rangle$,
$\langle C_L\rangle$, and $\langle C_U\rangle$ look very similar to
those shown for the RTI-stable case in Fig. \ref{fig:rthsheights}
under this identification.  As can be seen in
Fig. \ref{fig:rthsheights}, the depth associated with $\langle
C_L\rangle$ (corresponding to $h_{C_U}$ for the RTI-unstable
simulation) is very close to $d_T$ (corresponding to $h_T$ for the
RTI-unstable simulation).  However, the depth associated with $\langle
C_U\rangle$ (corresponding to $h_{C_L}$ for the RTI-unstable
simulations) is small in comparison to $d_T$.

\begin{figure}
\includegraphics[width=0.45\textwidth]{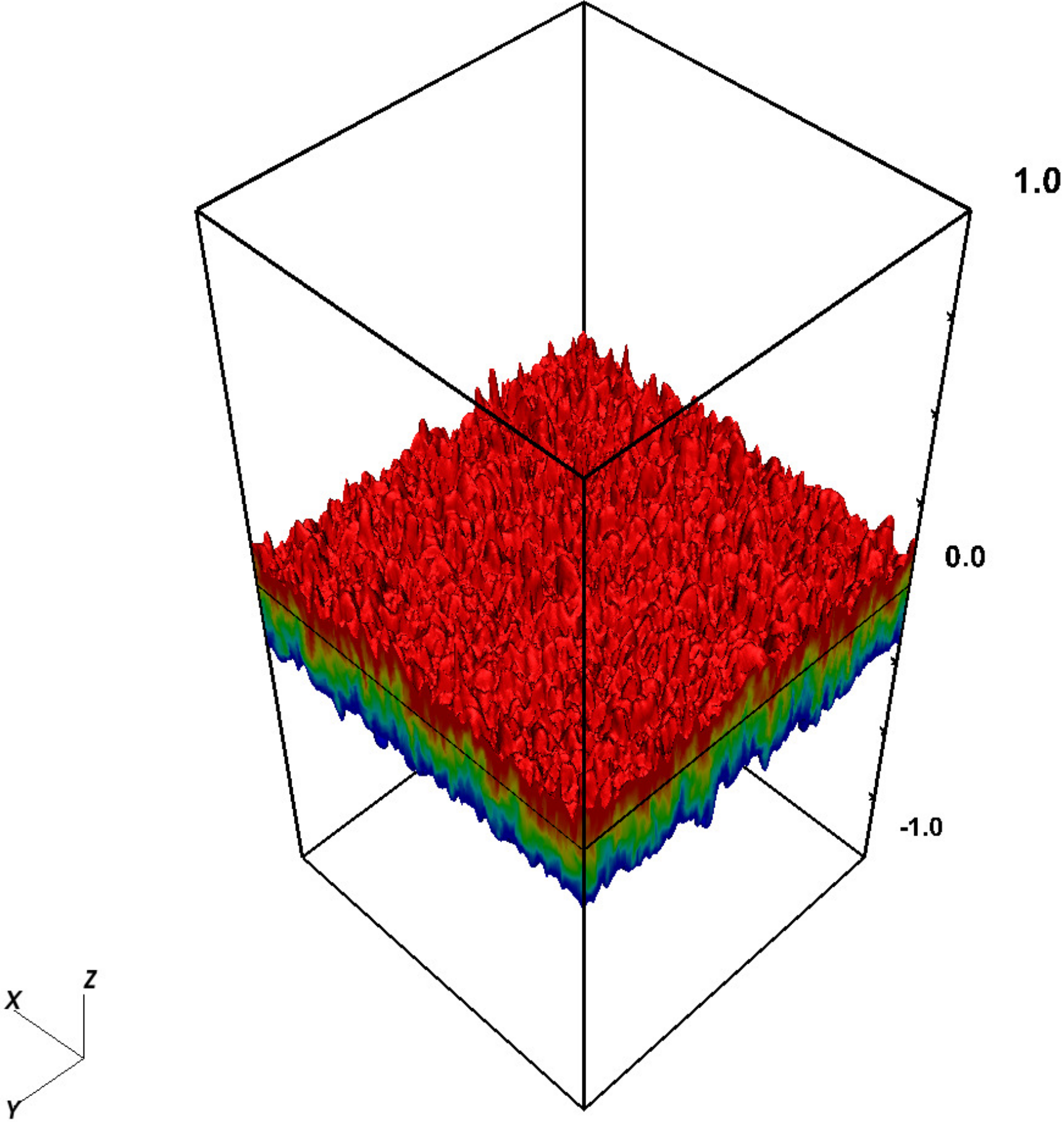}
\includegraphics[width=0.45\textwidth]{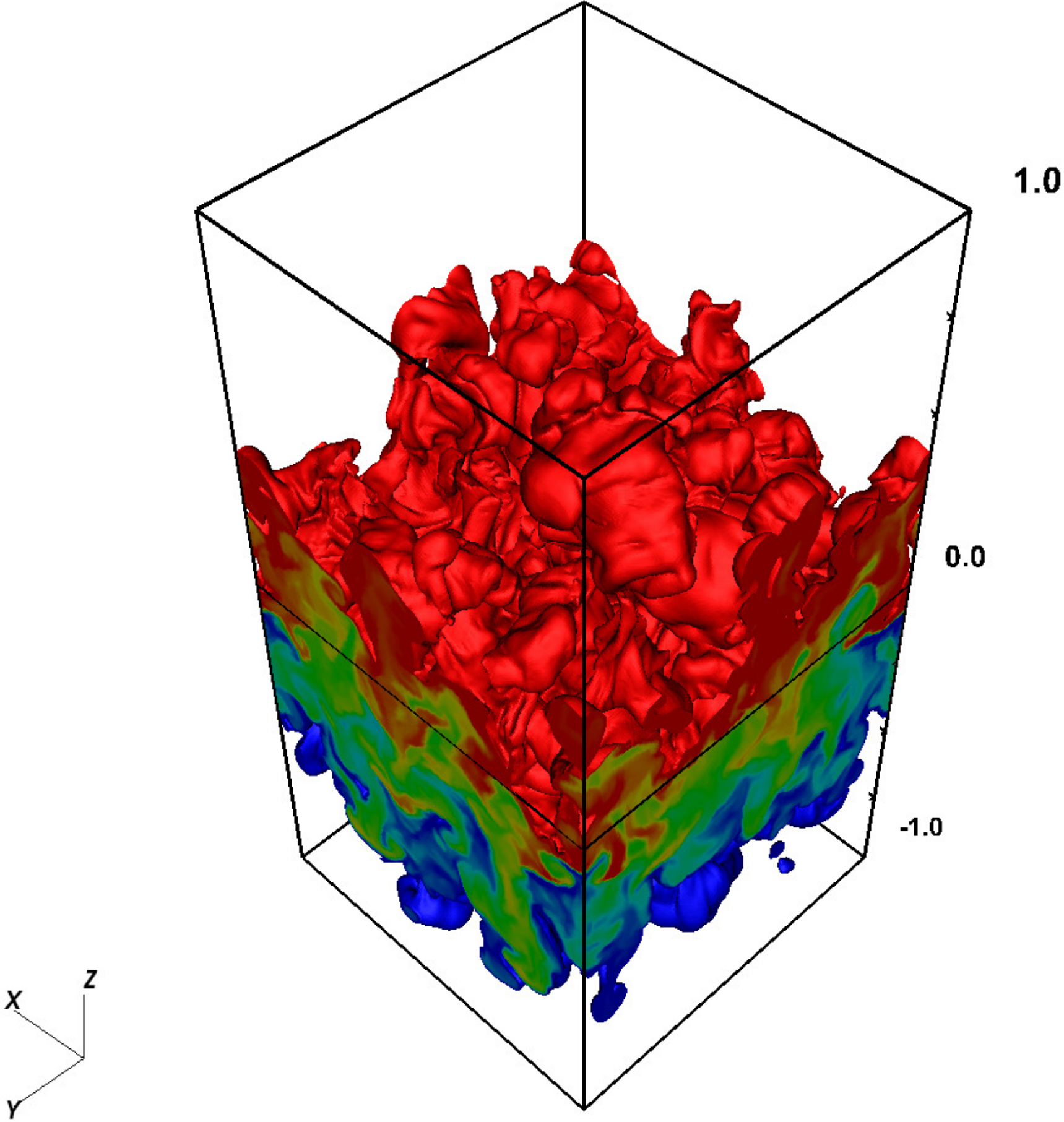}
\caption{Isosurfaces of temperature at $T=95$ and $T=38$ for the high
  resolution RTI-unstable simulation with an initially horizontal
magnetic field and with $\omega_c/\omega_\mathrm{RTI}=0.09$ (H2HR in
Table \ref{tab:cond}).  The upper panel is at time
$t/\sqrt{L/Ag}\approx 1.8$, and the lower panel is at time
$t/\sqrt{L/Ag}\approx 4.2$.  Temperature contours are also shown on
the faces of the simulation domain.  At early times, conduction is
more important than the RTI, and the temperature contours on the faces
of the simulation domain shown in the upper panel show that the
temperature mixing layer has diffused substantially.  The perturbation
on the $T=38$ surface are partly due to the tops of the RTI bubbles
which are growing despite the diffusion.  These bubbles soon emerge
from the diffusion region, as seen in the lower panel.}
\label{fig:rthucontours}
\end{figure}

In Fig. \ref{fig:rthucontours} we plot isosurfaces of temperature for
the RTI-unstable simulation with $\omega_c/\omega_\mathrm{RTI}=0.09$
at two times.  The upper panel shows the temperature at early times
when the RTI is just about to start setting the height, whereas the
lower panel shows the temperature at late times when the RTI is more
important than conduction.  The presence of some vertical conduction
is clear when comparing the upper panel of Fig. \ref{fig:rthucontours}
to the upper panel of Fig. \ref{fig:rtcontours}, which plots
isosurfaces of temperature at the same time for a run without any
explicit conduction.  The small scale perturbations on the surface of
the $T=38$ surface are due to a combination of uneven vertical
conduction -- as the vertical conduction depends on the size of the
initial perturbation -- as well as the tops of RTI bubbles which are
growing despite the diffusion.  The bubbles eventually grow faster
than the temperature mixing layer diffuses, and the dynamics become
dominated by the RTI (as seen in the lower panel).  However,
conduction is still important at late times, as the bubbles do not
rise as quickly as in simulations with no explicit conduction (see
Fig. \ref{fig:horunstable}).  This can also be seen in the lower panel
of Fig. \ref{fig:rthucontours} -- the RTI bubbles are much wider than
the RTI bubbles in our simulation with no explicit conductivity
(Fig. \ref{fig:rtcontours}).

We find that the presence of the MTI does not significantly alter the
RTI dynamics.  If we allow our simulations to run long enough that the
RTI bubbles hit the top and bottom walls, we begin to see convection.
If the conduction time is shorter than the buoyancy time, then this
convection is effectively driven by the MTI \citep{mcc11}.  However,
even without conduction, the simulations are convectively unstable at
late times, so the MTI does not introduce qualitatively different
dynamics into the problem.

There is significant growth in magnetic energy for the RTI-unstable
simulations (see Fig. \ref{fig:ME}).  Just as the magnetic energy
grows exponentially in the RTI-unstable simulation without any
explicit conductivity (solid line), the magnetic energy also grows
exponentially in the RTI-unstable simulation with an initially
horizontal magnetic field (dashed line).  The growth rates are similar
for both cases, although the growth in the simulation with anisotropic
conduction is somewhat slower (i.e., the exponential growth starts
later).  This is likely because bubbles rise more slowly in the
simulations with anisotropic conductivity.  By the end of both
simulations, the magnetic energy has amplified by about $10^3$.

\subsection{Vertical magnetic field}\label{sec:vertB}

We will now describe simulations initialized with a vertical magnetic
field.  Simulations with $\rho_+<\rho_-$ are still RTI-stable, but are
now HBI-unstable, yielding new dynamics.  The simulations with
$\rho_+>\rho_-$ are HBI-stable and RTI-unstable and are similar to the
RTI-unstable simulations with horizontal fields.  These simulations
are not initially in equilibrium, as there is a vertical temperature
discontinuity and vertical heat conduction.  Physically, the
simulations should be viewed as probing length scales much longer than
the initial width of the spread in temperature.  On these long length
scales, the temperature is initially very close to discontinuous.

\subsubsection{RTI-stable, HBI-unstable ($\rho_+<\rho_-$)}\label{sec:vertstable}

\begin{figure}
\includegraphics[width=0.45\textwidth]{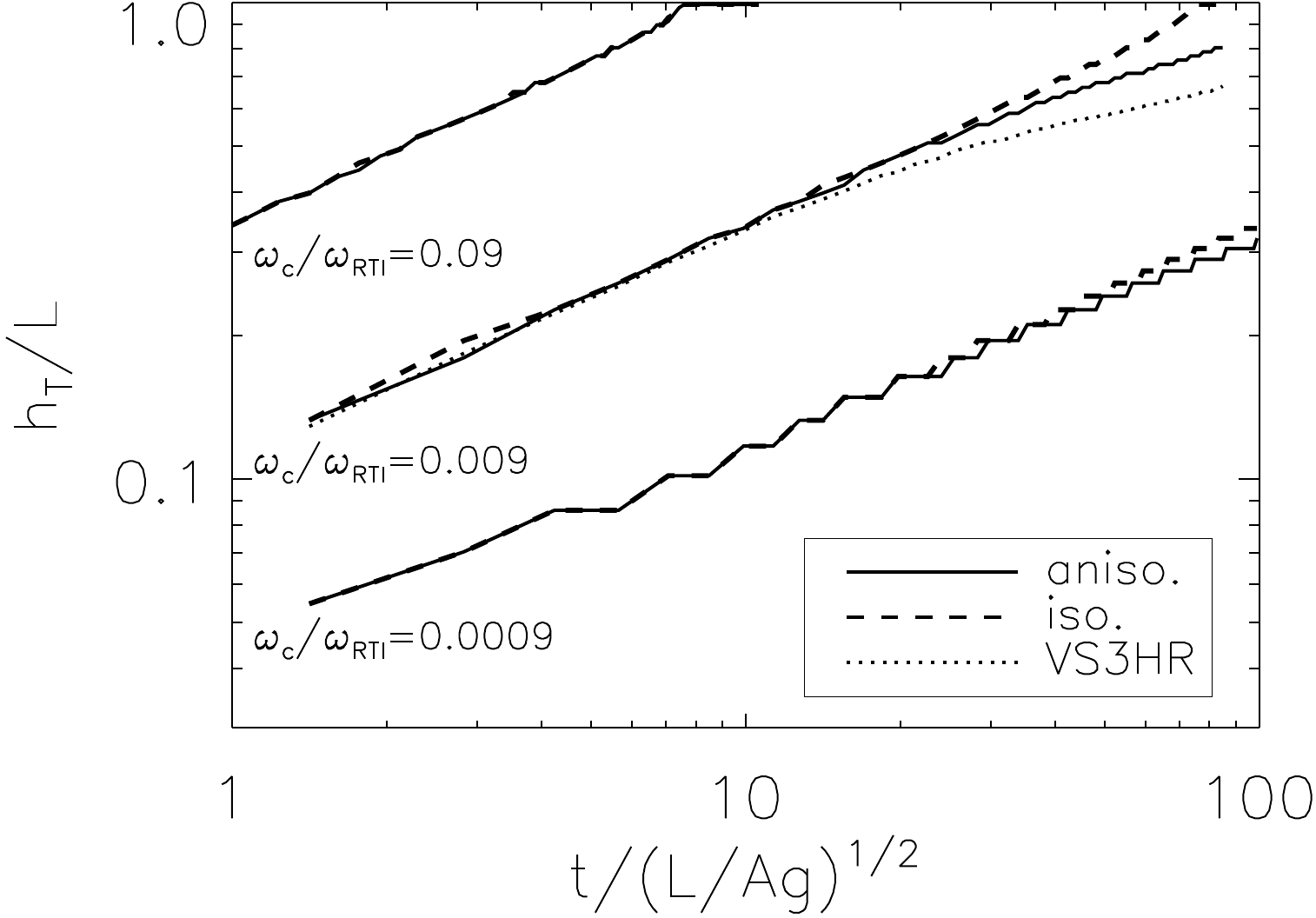}
\caption{The height of the temperature mixing layer (defined in
  \S \ref{sec:cond}) as a function of time for our RTI-stable
  simulations with vertical initial magnetic fields.  The solid lines
  show the results for simulations with anisotropic conductivity,
  whereas dashed lines show the results for simulations with isotropic
  conductivity.  The top pair of lines, which are overlapping, are for
  $\omega_c/\omega_\mathrm{RTI}=0.09$ (VS2 and VS2I in
  Table \ref{tab:cond}).  The middle set of three lines all correspond
  to $\omega_c/\omega_\mathrm{RTI}=0.009$.  The solid line has
  anisotropic conductivity (VS3), the dashed line has isotropic
  conductivity (VS3I), and the dotted line has anisotropic
  conductivity, but is at higher resolution (VS3HR).  The bottom two
  lines are for $\omega_c/\omega_\mathrm{RTI}=0.0009$.  For the
  simulations with relatively small anisotropic conductivities, the
  HBI becomes non-linear before the temperature mixing layer reaches
  the upper boundary of the simulation.  This results in deviations
  between the heights from the simulations with isotropic and
  anisotropic conductivity.  The HBI is more effective at diminishing
  conduction in the high resolution simulation.}
\label{fig:vertstable}
\end{figure}

We will first investigate the RTI-stable, HBI-unstable case.  These
runs simulate the discontinuous limit of the HBI.  The presence of the
HBI in our simulations can be inferred from the height of the
temperature mixing layer as a function of time, as shown in
Fig. \ref{fig:vertstable}.  The solid (dashed) lines correspond to
simulations with anisotropic (isotropic) conductivity.  The dotted
line shows the results of our high resolution simulation (VS3HR in
Table \ref{tab:cond}).  We find that $h_T$ increases as $\sqrt{\chi
t}$ for the simulations with isotropic conductivity.  For the
relatively high conductivity of $\omega_c/\omega_\mathrm{RTI}=0.09$,
the height of the temperature mixing layer is the same for the
simulation with anisotropic conductivity (VS2) as for the simulation
with isotropic conductivity (VS2I).  This is because the HBI has not
had enough time to modify the conduction across the entire simulation
domain before the temperature mixing layer hits the upper boundary.
However, for simulations with $\omega_c/\omega_\mathrm{RTI}=0.009$ or
$0.0009$ the HBI is able to partially inhibit conduction, slowing the
growth of the height of the temperature mixing layer.

The HBI saturates by reorienting vertical field lines to a more
horizontal geometry \citep{pq08}.  Horizontal fields are produced by
strong horizontal flows \citep{mcc11}.  This reduces the amount of
vertical heat conduction, slowing the growth of the temperature mixing
layer.  In the rapid conduction limit
($\omega_c\gtsimeq\omega_\mathrm{buoy}$), the saturation time-scale is
several times the buoyancy time-scale,
$\omega_\mathrm{buoy}=\sqrt{gd\log T/dz}$.  Conversely, if conduction
is slow ($\omega_c<\omega_\mathrm{buoy}$), the saturation time-scale is
several times the conduction time-scale.  The buoyancy frequency in our
simulations is initially very high because of the temperature
discontinuity, so the HBI initially grows on the conduction time-scale.
Thus, the HBI initially grows more slowly in runs with lower
conductivity than in runs with higher conductivity.  This is why the
HBI has had a more prominent effect on the height in the simulations
with $\omega_c/\omega_\mathrm{RTI}=0.009$ than in the simulations with
$\omega_c/\omega_\mathrm{RTI}=0.0009$ (see Fig. \ref{fig:vertstable}).  We
find that our high resolution simulation exhibits the slowest increase
in height of all the HBI-unstable simulations.  Due to increased
resolution, the field is able to become more horizontal than in the
simulations with our fiducial resolution, thereby decreasing the
effective vertical conductivity.

Previous simulations of the HBI show that although the field lines
never become horizontal, the angle they form with the horizontal
decreases as $\sim (t/t_\mathrm{buoy})^{-0.85}$ \citep{mcc11}, where
$t_\mathrm{buoy}\sim\omega_\mathrm{buoy}^{-1}$.  To measure the average
orientation of the magnetic field relative to the horizontal, we
define $\langle\hat{b}_z^2\rangle$ to be
\begin{equation}
\left<\hat{b}_z^2\right>=\left<\frac{B_z^2}{\boldsymbol{B}\boldsymbol{\cdot}\boldsymbol{B}}\right>,
\end{equation}
where $\langle\cdot\rangle$ denotes horizontal average, as above.  For
simulations with initially vertical magnetic fields,
$\langle\hat{b}_z^2\rangle$ is initially equal to one everywhere.

\begin{figure}
\includegraphics[width=0.45\textwidth]{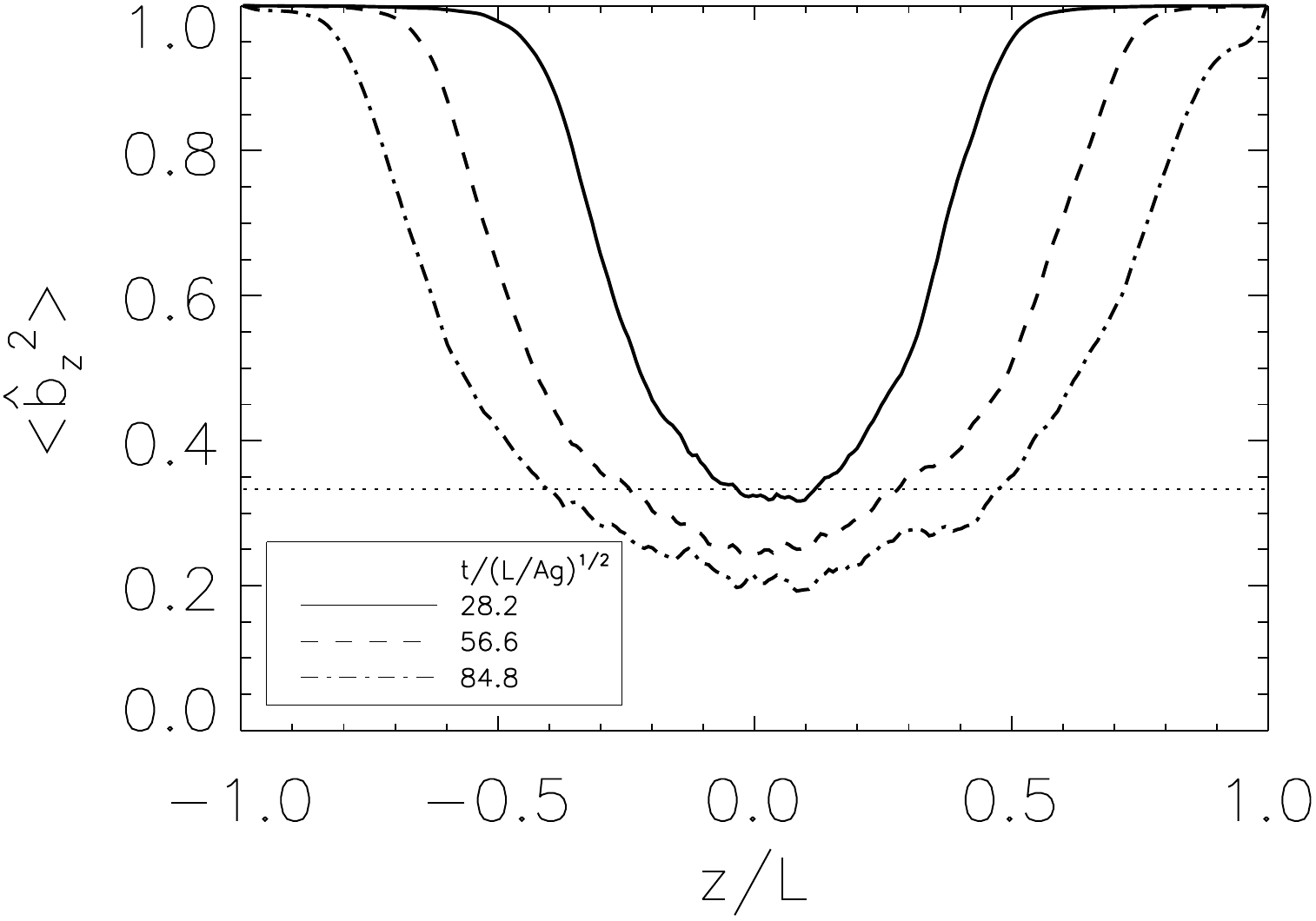}
\caption{$\langle \hat{b}_z^2\rangle$ as a function of height at
  various times for our high resolution simulation with
  $\omega_c/\omega_\mathrm{RTI}=0.009$ (VS3 in Table \ref{tab:cond}).
  We plot $\langle\hat{b}_z^2\rangle$ at $t/\sqrt{L/Ag}=28.2$ (solid
  line), $56.6$ (dashed line), and $84.8$ (dot-dashed line).  The
  horizontal dotted line shows $\langle\hat{b}_z^2\rangle=1/3$, which
  corresponds to an isotropic magnetic field.  Due to the action of
  the HBI, the magnetic field becomes increasingly horizontal as the
  simulation progresses in time, especially near the centre of the
  domain where the buoyancy time is the shortest.}
\label{fig:bz2}
\end{figure}

In Fig. \ref{fig:bz2} we plot $\langle\hat{b}_z^2\rangle$ as a
function of height for our high resolution simulation with
$\omega_c/\omega_\mathrm{RTI}=0.009$ (VS3HR in Table \ref{tab:cond}).
If thermal conduction is isotropic, $\langle\hat{b}_z^2\rangle$
remains close to one for the entire simulation.  Figure \ref{fig:bz2}
shows, however, that $\langle\hat{b}_z^2\rangle$ decreases due to the
HBI.  We know that the decrease in $\langle\hat{b}_z^2\rangle$ is not
due to random mixing, as $\langle\hat{b}_z^2\rangle$ decreases to
below the isotropic value of $1/3$.  There is also an asymmetry
between $\langle\hat{b}_z^2\rangle$ for $z<0$ and for $z>0$.  Because
the pressure force pushes the interface upwards
(see \S \ref{sec:horstable}), the HBI acts somewhat more strongly
for $z>0$ than for $z<0$, causing $\langle\hat{b}_z^2\rangle$ to be
smaller for $z>0$ than for $z<0$.

We can estimate how the change in orientation of the magnetic field
affects the evolution of $h_T$.  In the HBI non-linear evolution, the
magnetic field orientation changes as
\begin{equation}
\langle \hat{b}_z^2\rangle\sim (t/t_{buoy})^{-1.7}.
\end{equation}
As seen in Fig. \ref{fig:bz2}, $\langle \hat{b}_z^2\rangle$ is a
strong function of height.  However, for simplicity we will only
consider $\langle \hat{b}_z^2\rangle$ at $z=0$ and only consider heat
conduction through the $z=0$ plane.  The energy equation implies
$dh_T/dt\sim \chi_\mathrm{eff}/h_T$, where
$\chi_\mathrm{eff}=\chi\langle\hat{b}_z^2\rangle$.  We assume $d \log
T/dz\sim 1/h_T$ and thus equate $t_{buoy}\sim\sqrt{h_T/g}$.  One can
then show
\begin{equation}
\frac{d h_T}{dt}\sim\frac{\chi}{g^{.85}}\frac{1}{h_T^{.15}t^{1.7}}.
\end{equation}
This implies that $h_T$ should converge to a finite value at long
times, specifically
\begin{equation}
h_T\sim \frac{\chi^{2/3}}{g^{1/3}}.
\end{equation}
In this calculation we have neglected terms of order unity.  It is
unclear if $h_T$ is actually converging to a finite value in our
simulations that show clear evidence for the HBI.  This is not
surprising as even our simulations that start with horizontal
magnetic fields conduct heat vertically (see \S \ref{sec:horstable}).
Put another way, finite resolution prevents
$\langle \hat{b}_z^2\rangle$ from decreasing below some minimum value.

In our RTI-stable but HBI-unstable simulations with initially vertical
magnetic fields, we find that the evolution of the heights associated
with our passive scalars, $h_{C_L}$ and $h_{C_U}$, are qualitatively
similar to the results from the RTI-stable simulations with horizontal
initial magnetic fields.  Furthermore, the heights are very similar
when comparing simulations with isotropic and anisotropic conduction.
This is expected at early times, as there is little mixing occurring
at the interface until the HBI becomes non-linear.  Even in the
non-linear phase it is not that surprising that the HBI does not
contribute significantly to mixing at the interface: as described
in \citet{mcc11}, the HBI does not excite large velocities during its
saturation.

\begin{figure}
\includegraphics[width=0.45\textwidth]{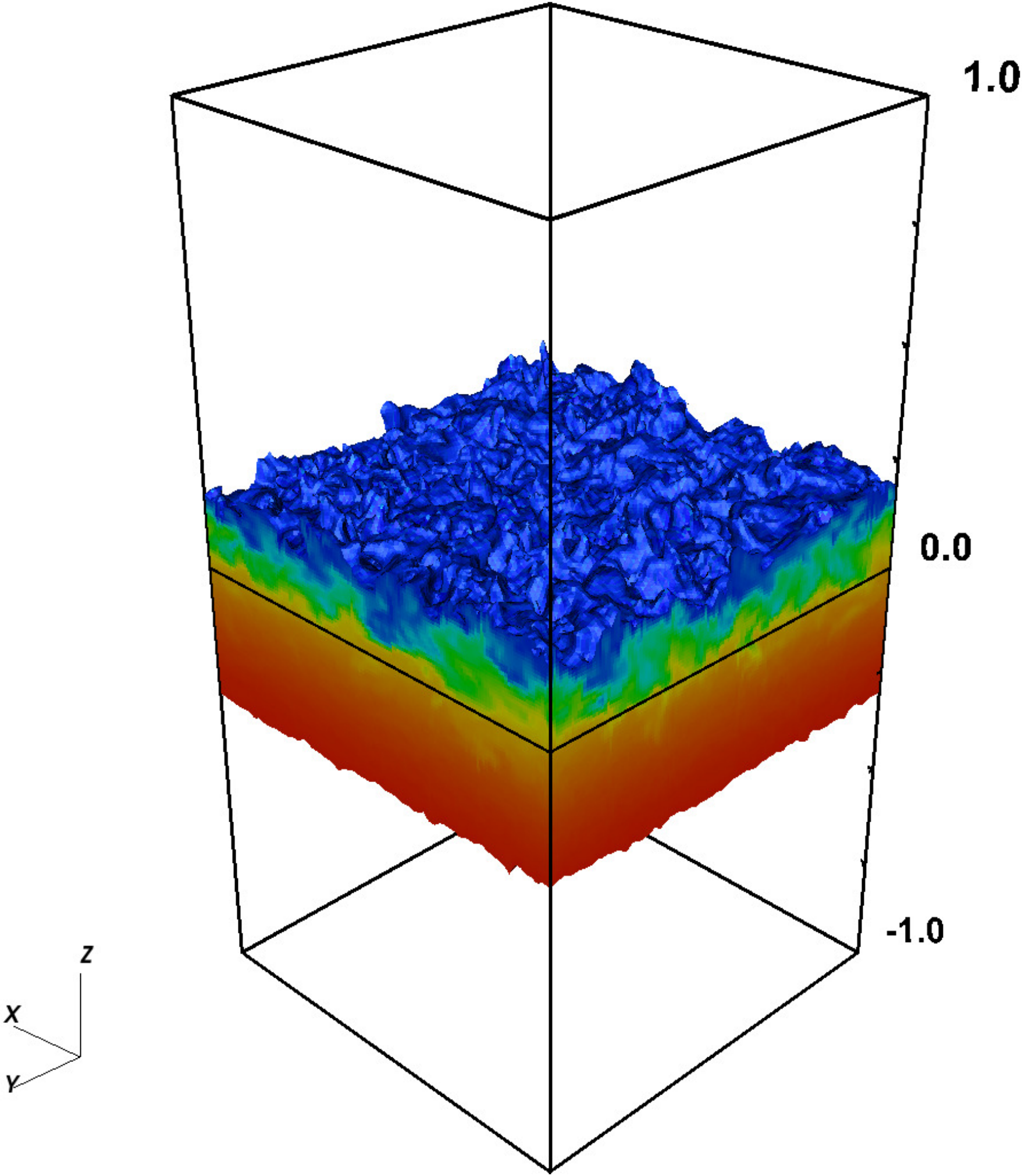}
\caption{Isosurfaces of fixed composition $C_U$ at $C_U=0.95$ and $C_U=0.05$ for the
high resolution RTI-stable, HBI-unstable simulation with
$\omega_c/\omega_\mathrm{RTI}=0.009$ (VS3HR in Table \ref{tab:cond})
at time $t/\sqrt{L/Ag}=71$.  Contours of $C_U$ are also shown on the
faces of the simulation domain.  For simulations with isotropic
viscosity $C_U$ has very little horizontal variation.  Thus, any
horizontal structure is due to the HBI.}
\label{fig:s0contours}
\end{figure}

In Fig. \ref{fig:s0contours} we plot isosurfaces of $C_U$ for the high
resolution HBI-unstable simulation with
$\omega_c/\omega_\mathrm{RTI}=0.009$ (VS3HR in Table \ref{tab:cond}).
The vertical extent of the composition mixing layer is much smaller
than the height of the temperature mixing layer, which is about $0.6$
(see Fig. \ref{fig:vertstable}).  There are also structures moving
horizontally due to the flow produced by the HBI.  This motion
produces the horizontal magnetic fields which are the saturated state
of the HBI.

We can see the magnetic field growth associated with the HBI in
Fig. \ref{fig:ME}.  Our high resolution RTI-stable simulation with a
vertical initial magnetic field (triple-dot-dashed line) experiences
linear growth in magnetic energy starting at $t/\sqrt{L/Ag}\approx 7$.
By the end of the simulation the magnetic energy has increased by a
factor of nine.  The simulation with an initially vertical magnetic
field but \textit{isotropic} conduction exhibits essentially no
magnetic field growth.  Thus, this linear magnetic field growth is due
to the HBI.  Because the HBI does not generate large velocities, the
growth is not as extreme as in the RTI-unstable simulations.

\subsubsection{RTI-unstable, HBI-stable ($\rho_+>\rho_-$)}
\label{sec:vertunstable}

\begin{figure}
\includegraphics[width=0.45\textwidth]{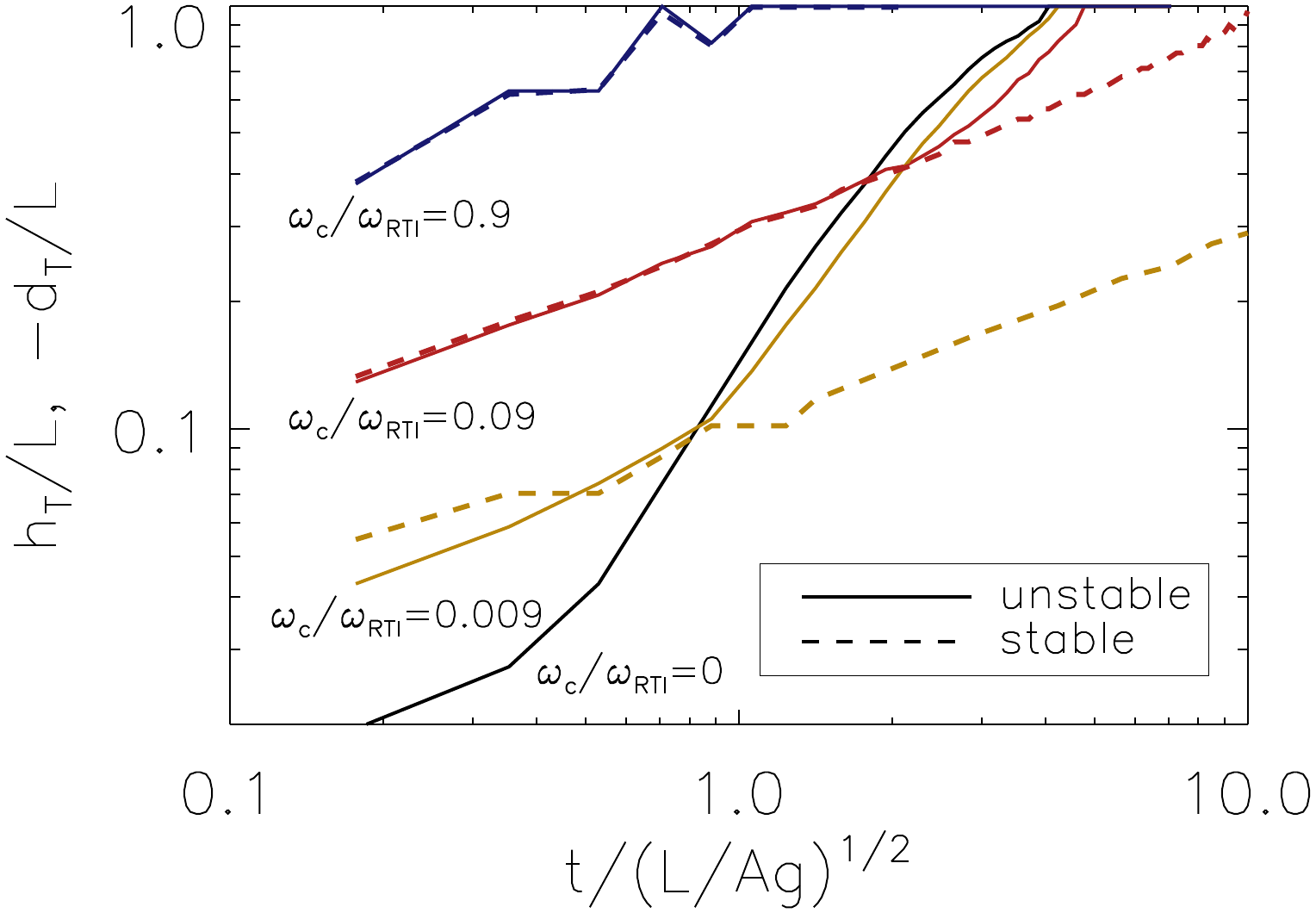}
\caption{Heights and depths of the temperature mixing layer (defined in
  \S \ref{sec:cond}) as a function of time for simulations with
  initially vertical magnetic fields, and with various
  $\omega_c/\omega_\mathrm{RTI}$.  The solid lines show $h_T$ for our
  simulations which are RTI-unstable, whereas the dotted lines show
  $-d_T$ for our simulations which are RTI-stable, and which
  have \textit{isotropic} conductivity.  The top pair of lines (in
  blue), which are overlapping, are for
  $\omega_c/\omega_\mathrm{RTI}=0.9$ (V1 and VS1I in
  Table \ref{tab:cond}).  The next pair of lines (in red) are for
  $\omega_c/\omega_\mathrm{RTI}=0.09$ (V2 and VS2I).  The last set of
  two lines (in yellow) correspond to
  $\omega_c/\omega_\mathrm{RTI}=0.009$ (V3 and VS3I).  The bottom solid line
  shows $h_T$ for our fiducial simulation without explicit
  conductivity (RT in Table \ref{tab:nocond}).  The heights for the
  RTI-unstable runs are about equal to the depths for the RTI-stable
  runs when conduction dominates over the RTI.  When the RTI becomes
  important, the heights in the RTI-unstable simulations increase in
  the same way as the height in the fiducial simulation.}
\label{fig:vertunstable}
\end{figure}

The RTI-unstable case with a vertical magnetic field is very similar
to the RTI-unstable case with an initially horizontal magnetic field
(\S \ref{sec:horunstable}).  We plot the height of the temperature
mixing layer as a function of time in Fig. \ref{fig:vertunstable}.
The results are very similar to the corresponding plot in the
RTI-unstable case with an initially horizontal magnetic field
(Fig. \ref{fig:horunstable}).  The results are similar because
initially both sets of simulations are dominated by vertical heat
conduction, but then transition to the RTI at later times.  The most
prominent difference is that in the initially vertical magnetic field
case, when the RTI becomes important the height becomes very close to
the height in the fiducial simulation.  This is because vertical heat
conduction does not impede the rise of RTI bubbles as much as
horizontal heat conduction, so diffusive effects slow the rise of RTI
bubbles less in the simulations with initially vertical magnetic
fields.  The RTI bubbles in simulations with initially vertical fields
are thinner than the RTI bubbles in simulations with initially
horizontal fields, as in our simulation without explicit conductivity
(see Fig. \ref{fig:rtcontours}).  As in the RTI-unstable simulations
with an initially horizontal magnetic field, the height in the high
resolution simulation grows slightly more slowly than at the fiducial
resolution.

The magnetic field growth for the RTI-unstable case with initially
vertical field is also similar to the case with initially horizontal
field (see Fig. \ref{fig:ME}).  As for all the RTI-unstable
simulations, the magnetic energy increases exponentially, and with
about the same growth rate as for the simulations with initially
horizontal magnetic fields.  However, the exponential growth of
magnetic energy begins much later (at $t/\sqrt{L/Ag}\approx 3$ rather
than $t\approx 0$) in the simulation with an initially vertical
magnetic field, and the magnetic energy only amplifies by about
$10^2$.  This is because there is not as much horizontal motion to
bend and amplify the initially vertical field lines.

\subsection{Skewed magnetic fields}

We have also run a set of simulations with initial magnetic fields
which make a 45 degree angle with the horizontal (runs A3 and AS2 in
Table \ref{tab:cond}).  These were compared to simulations with
initially vertical magnetic fields, but with anisotropic thermal
conduction with a magnitude one half as large (so the initial vertical
heat conduction, proportional to $\chi\langle\hat{b}_z^2\rangle$, was
kept constant).  The RTI-stable simulation with an initially skewed
magnetic field (AS2) was run with a conductivity of
$\omega_c/\omega_\mathrm{RTI}=0.18$, and is very similar to the
RTI-stable simulation with an initially vertical magnetic field, but
with $\omega_c/\omega_\mathrm{RTI}=0.09$ (VS2).  We did not run the
simulation with a sufficiently low conductivity to see the effects of
the HBI, which are presumably not as strong for the simulation with
initially skewed magnetic fields.

For RTI-unstable simulations, the simulation with an initially skewed
magnetic field and $\omega_c/\omega_\mathrm{RTI}=0.018$ (A3), and the
simulation with an initially vertical magnetic field and
$\omega_c/\omega_\mathrm{RTI}=0.009$ (V3) have the same height
initially, when the temperature mixing layer is primarily expanding
due to diffusion.  However, when the RTI becomes dominant, the height
for the simulation with an initially skewed magnetic field is somewhat
lower than the height for the simulation with the initially vertical
magnetic field.  This is because in the RTI-dominated regime, the
magnitude of the diffusivity matters more than the direction of the
initial field, as the magnetic field becomes isotropized.  The
simulation with an initially skewed magnetic field has
$\omega_c/\omega_\mathrm{RTI}$ twice as large as the simulation with
an initially vertical magnetic field, so this simulation is more
dissipative.  This causes a slower increase in height.

In physical systems, the magnetic field will never be perfectly
aligned or perpendicular to the contact discontinuity.  However, these
results show that if the magnetic field makes a moderate angle with
the contact discontinuity, the behavior is very similar to a system
with a vertical magnetic field (see \S \ref{sec:vertB}), but with a
smaller conductivity.  If the magnetic field makes a small angle with
the contact discontinuity, then the evolution will likely be as
described by the horizontal magnetic field simulations
(see \S \ref{sec:horB}), which did have some vertical temperature
diffusion.

\subsection{Additional physical effects}\label{sec:anisovisc}

Thermal conduction is anisotropic when the electron gyroradius is much
smaller than the electron mean free path.  When this is true, the ion
gyroradius is often also smaller than the ion mean free path.  In this
case, viscosity also acts predominantly along magnetic field lines.
Because momentum is carried by the ions, the viscosity is smaller than
the thermal diffusivity by a factor of about 50 or 100 if $\mu=0.5$ or
$1$, respectively.  We carried out simulations implementing
anisotropic viscosity \citep[as described in][]{par12} for contact
discontinuities where are RTI-unstable.  We found that the height of
the temperature mixing layer is not significantly affected by
inclusion of anisotropic viscosity.  This is probably because the
viscosity is so much smaller than the thermal diffusivity.  For
simulations with an initially horizontal magnetic field, the RTI
bubbles were more elongated in the direction of the initial magnetic
field than in the simulations with only anisotropic conductivity and
numerical viscosity \citep[see also][]{ds09}.

\section{Summary and Discussion}\label{sec:conclusion}

In a dilute magnetized plasma, thermal conduction is anisotropic; this
can lead to qualitative changes in the behavior of the plasma.  We
have investigated the evolution of RTI-stable and RTI-unstable contact
discontinuities with anisotropic thermal conduction.  The linear
problem is only well-posed for a horizontal initial magnetic field.
In this case, there is no change to the RTI dispersion relation in the
local, Boussinesq limit.  In this limit, the classical RTI
perturbations are isothermal so they are unaffected by anisotropic
conduction (eqn. \ref{eq:isothermal}).  More generally, compressible
perturbations are affected by anisotropic conduction
(eqn. \ref{eq:RTfull}).  If the conduction time-scale is short compared
to the growth time of the RTI, perturbations are isothermal and
grow faster than the adiabatic perturbations of the classical RTI.
The enhancement in the growth rate is modest, but is a function of the
temperature contrast across the contact discontinuity and the
compressibility of the mode (see Fig. \ref{fig:omega}).

We have run a number of fully non-linear simulations of contact
discontinuities that are either RTI-stable or RTI-unstable, and with
several initial magnetic field geometries (see Fig. \ref{fig:chart}).
We focus on initially horizontal and initially vertical magnetic
fields -- these two choices of the magnetic field orientation bracket
the range of dynamics associated with contact discontinuities in
dilute plasmas.  We primarily used the heights of the temperature and
composition mixing layers as diagnostics for the evolution of the
initial temperature and density discontinuity.  Interestingly, in our
high resolution simulation with no explicit conduction (i.e.,
simulating the classic RTI) the height increases linearly with time in
the non-linear phase, rather than quadratically (see
Fig. \ref{fig:rtheight}).  This result is independent of the exact
definition of height we use, or when we assume self-similar non-linear
growth begins.

In a dilute plasma, the non-linear evolution of a contact discontinuity
can be described as a combination of three processes: vertical
temperature diffusion, the RTI, and the HBI.  For the simulations of
RTI-stable contact discontinuities with anisotropic conduction,
vertical temperature diffusion is most important.  This is true even
in our simulations with horizontal initial magnetic fields.  The
diffusion in this case occurs because we initially perturb the
vertical velocity, and subsequent numerical mixing across the
interface.  Thus, both in simulations with vertical initial magnetic
fields and horizontal initial magnetic fields, the height of the
temperature mixing layer increases due to diffusion.  Although there
is significant temperature diffusion, there is not much mixing at the
interface in the absence of the RTI, so the height of the composition
mixing layer remains small (see Fig. \ref{fig:rthstable}).

An RTI-stable contact discontinuity is unstable to the HBI in the
presence of vertical magnetic fields; the rate of vertical temperature
diffusion across the contact discontinuity slows significantly due to
the instability (see Fig. \ref{fig:vertstable}).  The HBI reorients
the vertical magnetic fields to a more horizontal geometry, impeding
further heat conduction.  We estimate that the HBI would cause a
contact discontinuity to spread to a finite width
$\sim\chi^{2/3}/g^{1/3}$.

Our simulations with RTI-unstable contact discontinuities initially
evolve in the same way as the simulations with RTI-stable contact
discontinuities -- at early times, the simulations are dominated by
temperature diffusion (see, e.g., Fig. \ref{fig:horunstable}).
However, RTI bubbles are forming and beginning to rise even as the
temperature discontinuity diffuses away.  The RTI bubbles rise faster
and overtake the temperature diffusion, leading to dynamics dominated
by the RTI at late times.  At this stage, anisotropic conductivity
does not play an especially important role.  The heights of the RTI
bubbles are similar in the simulations with and without anisotropic
conductivity.  However, the RTI bubbles are a little lower in the
simulations with anisotropic conductivity, due to increased
dissipation in the system.  This is especially true for the
simulations with initially horizontal magnetic fields.  In this case,
the RTI bubbles leak heat to their sides, becoming less buoyant.

We now discuss astrophysical contexts in which our results may be
applicable, specifically, supernova remnants, emerging magnetic flux in
the solar corona, and cold fronts in galaxy clusters.

\citet{tb06} discuss the effects of anisotropic conduction in a
supernova remnant.  They simulated an entire supernova remnant, but
unlike \citet{che92} did not find any RTI.  Surprisingly, it does not
seem that there is significant temperature conduction across their
contact discontinuity, as the remnant maintains its spherical shape,
instead of becoming more cylindrical as one would expect with their
initial magnetic field.

In supernova remnants, the contact discontinuity feels an effective
gravitational force due to its deceleration into the ambient medium.
The magnitude of the acceleration is decreasing with time as
$g_\mathrm{eff}\sim t^{-8/5}$ \citep[e.g.,][]{sed46}.  Thus, one might expect the height of
RTI bubbles to increase as $g_\mathrm{eff} t^2\sim t^{2/5}$.  At late
times this is slower than the diffusive growth of the height of the
temperature mixing layer, which grows as $\sqrt{\chi t}$.  As an
example, assume a supernova remnant with an electron temperature of
$T=10^8\;\mathrm{K}$ and number density $n=4\;\mathrm{cm}^{-3}$ is
expanding into a medium with $n=1\;\mathrm{cm}^3$.  Then we have that
the height of the temperature mixing layer due to temperature
diffusion after a Sedov time $\sim 100\;\mathrm{yr}$ is
\begin{equation}
\sqrt{\chi t}\sim 3\;\mathrm{pc}\;\left(\frac{t}{100\;\mathrm{yr}}\right)^{1/2}.
\end{equation}
After $\sim 100\;\mathrm{yr}$ the remnant would expand to a radius of
about $r\sim 2\;\mathrm{pc}$, so $g_\mathrm{eff}\sim r/t^2\sim
0.6\;\mathrm{cm}\;\mathrm{s}^{-2}$.  We find that the height of the
highest RTI bubble after $100\;\mathrm{yr}$ would be about
\begin{equation}
g_\mathrm{eff} t^2\sim 2\;\mathrm{pc}\;\left(\frac{t}{100\;\mathrm{yr}}\right)^{2/5}.
\end{equation}
This suggests that diffusion and RTI dynamics are about equally
important after a Sedov time, although diffusion would likely dominate
at later times.  Thus, a global simulation of an entire supernova
remnant would be necessary to accurately determine how anisotropic
conduction affects the contact discontinuity.

\citet{ber11} recently reported observations of hot, dilute
solar prominences.  They interpret the observations as showing the RTI
on the surface of the bubble of hot plasma, as described in numerical
simulations \citep[e.g.,][]{iso05}.  They report temperatures of the
hot plasma to be $T\approx 10^6\;\textrm{K}$, which, using $n\approx
10^9\;\textrm{cm}^{-3}$, gives a collision frequency of $\nu\approx
10^2\;\textrm{Hz}$.  This is much smaller than the electron
gyrofrequency $\omega_\mathrm{ce}\approx 10^7
(B/1\;\mathrm{G})\;\mathrm{Hz}$, indicating the heat conduction is
primarily along magnetic field lines.  Furthermore, such a plasma would
have a parallel heat diffusivity of $\approx
10^{16}\;\mathrm{cm}^2\;\mathrm{s}^{-1}$ \citep{spitzer62}, so the diffusion
time across a prominence which has a size $\approx 10^9\;\mathrm{cm}$
is about $100\;\mathrm{s}$.  This is much shorter than the evolution
time for the prominence, which is on order an hour.  Since conduction
becomes increasingly important at smaller length scales, this shows
that anisotropic conduction should be important for the RTI discussed
in
\citet{ber11}.

The magnetic field in prominences is believed to be parallel to the
interface of the prominence.  If some field lines were perpendicular
to the interface, then heat conduction would quickly smear out the
interface (see \S \ref{sec:vertunstable}).  However, the observations
show that the prominences hold their shape on time-scales of an hour.
Thus, we can take the magnetic fields to be primarily along the
interface.  Our linear results in \S \ref{sec:review} show that the
growth of the RTI should be faster than would be predicted without
anisotropic conduction, though how much faster depends on the
compressibility of the modes.  In the non-linear regime
(\S \ref{sec:horunstable}), we expect a superposition of spreading of
the interface due to perpendicular heat diffusion and a somewhat
slower RTI growth.  Because perpendicular diffusion is not very
important (the perpendicular diffusivity is smaller by a factor of
$(\nu/\omega_\mathrm{ce})^2$), we expect the non-linear behavior to be
similar to the RTI without conduction.

Our calculations are not ideally suited for comparison with solar
prominences as we have assumed that the magnetic fields are
dynamically unimportant, which is not the case for solar prominences.
Strong magnetic fields inhibit motions which bend the field lines,
instead favoring interchange-like motions (SG07).  Anisotropic
viscosity along field lines also would inhibit motion along field
lines \citep{ds09}.

\citet{pq08} and \citet{bir10} suggested that thermal diffusion across the contact discontinuities found
at cold fronts in galaxy clusters could be suppressed by the HBI.
In \S \ref{sec:vertstable} we determined that the HBI will cause the
contact discontinuity to diffuse to a finite width
$\sim\chi^{2/3}/g^{1/3}$.  \citet{bir10} simulate shocks propagating
through a galaxy cluster which merge producing a cold front.  Using
their cluster parameters for a front at $200\;\mathrm{kpc}$, we find
that $\chi\approx 3\times 10^{30}\;\mathrm{cm}^2\;\mathrm{s}^{-1}$, and
$g\approx 10^{-8}\;\mathrm{cm}\;\mathrm{s}^{-2}$.  This gives the
saturation width of the contact discontinuity to be $\sim
30\;\mathrm{kpc}$, which is somewhat larger than the Chandra upper
limit of $10\;\mathrm{kpc}$.  The saturation width of the cold front
observed in the galaxy cluster A3667 \citep{vik01} can be estimated to also
be $\sim 40\;\mathrm{kpc}$ -- much larger than the observed width
$\leq 5\;\mathrm{kpc}$.  Furthermore, our numerical results
in \S \ref{sec:vertstable} suggests that the saturation width is
likely a few times $\chi^{2/3}/g^{1/3}$.  These results suggest that
additional effects not included in our simple model might be required
to explain the narrow widths of observed cold fronts.

\appendix
\section{Calibration of Linear Growth Rates using Numerical Simulations}

We have tried to confirm the growth rates derived
in \S \ref{sec:review} through numerical simulations.  Our numerical
methodology is summarized in \S \ref{sec:numerics}.  There are two
difficulties that make accurate comparison between linear theory and
the numerical results impossible.  The first difficulty is that the
fastest growing modes are those on the smallest length scale allowed
in the simulation.  The second difficulty is that viscous effects will
act to slow the growth of the instability.  Although we do not
implement an explicit viscosity in our numerical simulations
(see \S \ref{sec:numerics}), there is numerical diffusion which adds
an effective viscosity to the problem.  This viscosity will damp out
any modes with wavelengths smaller than a cutoff wavelength, which is
the width of a few grid points -- we will refer to this cutoff
wavelength as the \textit{viscous length scale}.  The fastest growing
modes will be those with a wavelength only slightly larger than this
viscous length scale.  Thus, these modes are substantially affected by
viscosity, and cannot be accurately compared to our calculations above
which are for the inviscid RTI problem.

A possible way to get around these difficulties would be to only
excite long wavelength modes which are not affected by viscosity.
However, rounding errors produce random perturbations at the smallest
scales in the simulation, and these random perturbations grow faster
than the long wavelength modes of interest (as their growth rates are
many times larger than the growth rates of the modes of interest).  If
we increase the amplitude of the initial perturbations, it will take
longer for the random noise to grow to the same size as the initial
perturbation.  Unfortunately, non-linear effects become important at
these higher amplitudes, so we still cannot accurately measure the
linear growth rate.

We have tried using the eigenfunctions from the calculation
in \S \ref{sec:review} as an initial perturbation, in hopes of seeing
some growth due to the initial perturbation before the simulation is
overwhelmed by the small scale modes.  Unfortunately, the cusp in the
perturbation vertical velocity and discontinuities in perturbation
horizontal velocity, density, and pressure at $z=0$ are not resolved
by the simulation.  Because of this, using the eigenfunctions as an
initial perturbation launches sound waves with amplitudes similar to
the amplitude of the initial perturbation.  These obfuscate the growth
of the mode of interest.  To summarize, we have sound waves launched
by the initial perturbation which dominate the dynamics at the
beginning of the simulation, and quickly growing small scale modes
which dominate later on in the simulation.  Between these two effects,
we are unable to measure the growth of the longer wavelength modes
which are relatively unaffected by viscosity.  Presumably these
difficulties could be remedied by using a resolved density profile
rather than a true discontinuity on the grid scale.  However, we
concluded that there was not sufficient motivation to do so given that
our numerical methods have been validated using many other tests (in
particular, see \citet{ps05} and \citet{sh07} for tests of the
anisotropic conduction methods).

\acknowledgments This material is based upon work supported by the National Science Foundation Graduate 
Research Fellowship under Grant No. DGE 1106400.  D. L. acknowledges
support from a Hertz Foundation Fellowship. I. J. P. and E. Q. are
supported in part by NASA Grant ATP09-0125, NSF-DOE Grant PHY-0812811,
and by the David and Lucille Packard Foundation.  We thank S. Balbus,
B. Brown, A. Schekochihin, G. Vasil, and E. Zweibel for useful
discussions and suggestions.  J. Luo and M. McCourt aided in figure
preparation.  Computing time was provided by the National Science
Foundation through the Teragrid resources located at the National
Center for Atmospheric Research.\\

\bibliographystyle{apj}

\bibliography{rtc}

\end{document}